\documentclass[namedreferences]{solarphysics}
\usepackage[optionalrh,solaromanenum]{spr-sola-addons} 
\usepackage{graphicx}                    
\usepackage{color}                       
\usepackage{url}                         
\usepackage{hyperref}

\begin{document}

\begin{article}

\begin{opening}

\title{Development of an Advanced Automated Method for Solar Filament Recognition and Its Scientific Application to a Solar
Cycle of MLSO H$\alpha$ Data }

%
\author{Q.~\surname{Hao}$^{1}$\sep
        C.~\surname{Fang}$^{1,2}$\sep
        P. F.~\surname{Chen}$^{1,2}$
       }

%
\runningauthor{Q. Hao, \textit{et al.}}
\runningtitle{An Advanced Automated Method for Solar Filament Recognition}

%
  \institute{$^{1}$ School of Astronomy and Space Science, Nanjing University, Nanjing 210093, China
                     email: \url{fangc@nju.edu.cn}\\
             $^{2}$ Key Laboratory for Modern Astronomy and Astrophysics (Nanjing University),
             Ministry of Education, Nanjing 210093, China}


\begin{abstract}

We developed a method to automatically detect and trace solar filaments in
H$\alpha$ full-disk images. The program is able not only to recognize
filaments and determine their properties, such as the position, the area, the
spine, and other relevant parameters, but also to trace the daily evolution of
the filaments. The program consists of three steps: First, preprocessing is
applied to correct the original images; Second, the Canny edge-detection method
is used to detect filaments; Third, filament properties are recognized
through the morphological operators. To test the algorithm, we applied it to
the observations from the Mauna Loa Solar Observatory (MLSO), and the program is
demonstrated to be robust and efficient. H$\alpha$ images obtained by MLSO from
1998 to 2009 are analyzed, and a butterfly diagram of filaments is
obtained. It shows that the latitudinal migration of solar filaments has three
trends in the Solar Cycle 23: The drift velocity was fast from 1998 to the
solar maximum; After the solar maximum, it became relatively slow. After 2006,
the migration became divergent, signifying the solar minimum. About 60\%
filaments with the latitudes larger than $50^{\circ}$ migrate towards the polar
regions with relatively high velocities, and the latitudinal migrating speeds
in the northern and the southern hemispheres do not differ significantly in the
Solar Cycle 23.
\end{abstract}

%

\end{opening}


\section{Introduction}

Solar filaments, called prominences when they appear above the solar limb, are
important magnetized structures containing cool and dense plasma embedded the
hot solar corona. Typically, a filament is 100 times cooler and denser than
its surrounding corona. They are particularly visible in H$\alpha$ observations,
where they often appear as elongated dark features with several barbs
\cite{1995Tandberg-Hanssen,2010Labrosse}. Filaments are always aligned with
photospheric magnetic-polarity inversion lines \cite{1998Martin} and are
located at a wide range of heliocentric latitudes.
This characteristic makes filaments suitable for tracing and
analyzing the solar magnetic fields
\cite{1972McIntosh,1994Mouradian,1998Minarovjech,1998Rusin}.
Moreover, filaments sometimes undergo large-scale instabilities, which
break their equilibria and lead to eruptions, so they are
often associated with flares and coronal mass ejections (CMEs)
\cite{2000Gilbert,2003Gopalswamy,2004Jing,2008Chen,2011Chen,2012Zhang}.
Therefore, both case study and statistical analysis of filaments are
important and significant.

With the rapid development of the telescopes, both time cadences and spatial
resolutions of the observations are becoming higher and higher. As a
consequence, we have to deal with a vast amount of data, and automated
detection is an efficient way to derive the features of interest in the
observations. In terms of solar filaments, a number of automated filament
detection methods and algorithms have been developed in the past decade. For
example, \inlinecite{2002Gao} combined the intensity threshold and region
growing methods in order to develop an algorithm to automatically detect the
growth and the disappearance of filaments. \inlinecite{2003Shih} adopted local
and global thresholding and employed morphological closing operations to
identify filaments. \inlinecite{2005Fuller} utilized morphological ``hit or
miss" transformation and calculated Euclidean distance to get the filament
spines.  \inlinecite{2005Bernasconi} developed an algorithm based on a
geometric method, which was recently updated by \inlinecite{Martens2012}, to
determine the filament chirality in addition to the locations, where they
confirmed the hemispheric rule of the filament chirality. Based on the Sobel
operator, \inlinecite{2005Qu} applied an adaptive threshing method to detect
and derive various parameters of filaments. \inlinecite{2010Wang} employed
morphological methods, while \inlinecite{2010Labrosse2} applied the Support
Vector Machine (SVM) method to detect EUVI 304 {\AA} prominence above the solar
limb. \inlinecite{2011Yuan} designed a cascading Hough circle to determine the
center location and the radius of the solar disk, and further to find the
filament spines based on graph theory.

In this paper, we present an efficient and versatile automated detecting and
tracing method for solar filaments. It is able not only to recognize
filaments, determine their features such as the position, the area, the spine,
and other relevant parameters, but also to trace the daily evolution of the
filaments. In Section~\ref{Preproc} we describe image preprocessing before
detecting filaments. The filament detection algorithm based on the Canny
edge-detection method and connected components process are given in
Section~\ref{Detection}. A detailed description of the feature recognition
algorithm is given in Section~\ref{Feature}. The tracing algorithm is explained
in Section~\ref{Tracing}. The performance of our program is described in
Section~\ref{Performance}. Finally, statistical results about the filament
latitudinal distribution based on the H$\alpha$ archive of Mauna Loa Solar
Observatory (MLSO) are presented in Section~\ref{Results} before the
conclusions are drawn in Section~\ref{Conclusion}.

\section{Preprocessing}
\label{Preproc}
The raw image preprocessing consists of five steps, which are explained one by
one in the following subsections.

\subsection{H$\alpha$ data acquisition and analysis}

The full-disk H$\alpha$ images that we processed are mainly downloaded from the
MLSO website (\url{http://mlso.hao.ucar.edu}). Each image has a size of
1024$\times$1024 pixels and is taken by the \textit{Polarimeter for Inner
Coronal Studies} (PICS). The pixel size of the image is
2.9$^{\prime\prime}$. The MLSO H$\alpha$ data archive provides two types of
images: one has a limb-darkening correction applied along with contrast
enhancement, the other is the raw data. Our program can process both the
``FITS" format and the web image formats such as ``GIF" and ``JPEG".

\subsection{Limb-Darkening Removal}

The limb-darkening effect, \textit{i.e.} the intensity drops towards the solar
limb, may cause false detections. We should remove it first. Some observatories
such as MLSO, also provide limb-darkening corrected images. A polynomial
fitting method of Keith Pierce \cite{2000asqu} is adopted to remove the
limb-darkening effect:

   \begin{equation}\label{ldr-equation}
   I_{\rm H\alpha}^{\rm cor}(\theta) = I_{\rm H\alpha}^{\rm raw}(\theta)/(1-u_{2}-v_{2}+u_{2}
   \cos\theta+v_{2}\cos^{2}\theta)\, ,
   \end{equation}

\noindent
where $I_{\rm H\alpha}^{\rm cor}(\theta)$ is the corrected intensity and
$I_{\rm H\alpha}^{\rm raw}(\theta)$ is the intensity in the raw images, $\theta$
is the angle between the local radial direction and the line of sight,
$u_{2}=0.88$ and $v_{2}=-0.23$ are the fitted constants for the H$\alpha$
wavelength at 6563 {\AA}.

\subsection{Solar Disk Extraction}

Since the solar disk is only a part of the entire image, surrounded by a large
part of the sky background, we need to remove the background in order to
process the solar disk only, which can reduce the processing time and the
storage space. The method for the disk extraction is simple: we just find the
left, the right, the top, and the bottom ranges of the solar disk. Then we get
the sub-image according to these ranges. An example is shown in
Figure~\ref{mlso_pre}(b).

\subsection{Top-hat Filter for Enhancement}

Morphological image processing is a type of processing in which the spatial
forms or the structures of the objects within an image are modified
\cite{1992Haralick,2001Pratt}. Erosion, dilation, opening (erosion
followed by dilation), and closing (dilation followed by erosion) are
the basic operators in the morphological concepts that have been extended to
work with gray-scale
images for image segmentation and enhancement. Sometimes we get the
images where the boundaries of filaments are not very clear. In
order to make more accurate segmentation of the filament structure,
it is necessary to enhance the image to increase
the intensity contrast between the filament and non-filament structures.
We use the morphological top-hat transformation to enhance the images.
The algorithm is composed of three steps:

\textbf{i)} To compute the morphological opening of the image with the top-hat
filtering and then to subtract the result from the original image;

\textbf{ii)} To compute the morphological closing of the image with the
bottom-hat filtering and then to subtract the result from the original image;

\textbf{iii)} To add the top-hat filtered image to the original image, and then
to subtract the result from the bottom-hat filtered image.

\begin{figure}    
   \centerline{\hspace*{0.0\textwidth}
               \includegraphics[width=0.59\textwidth,clip=]{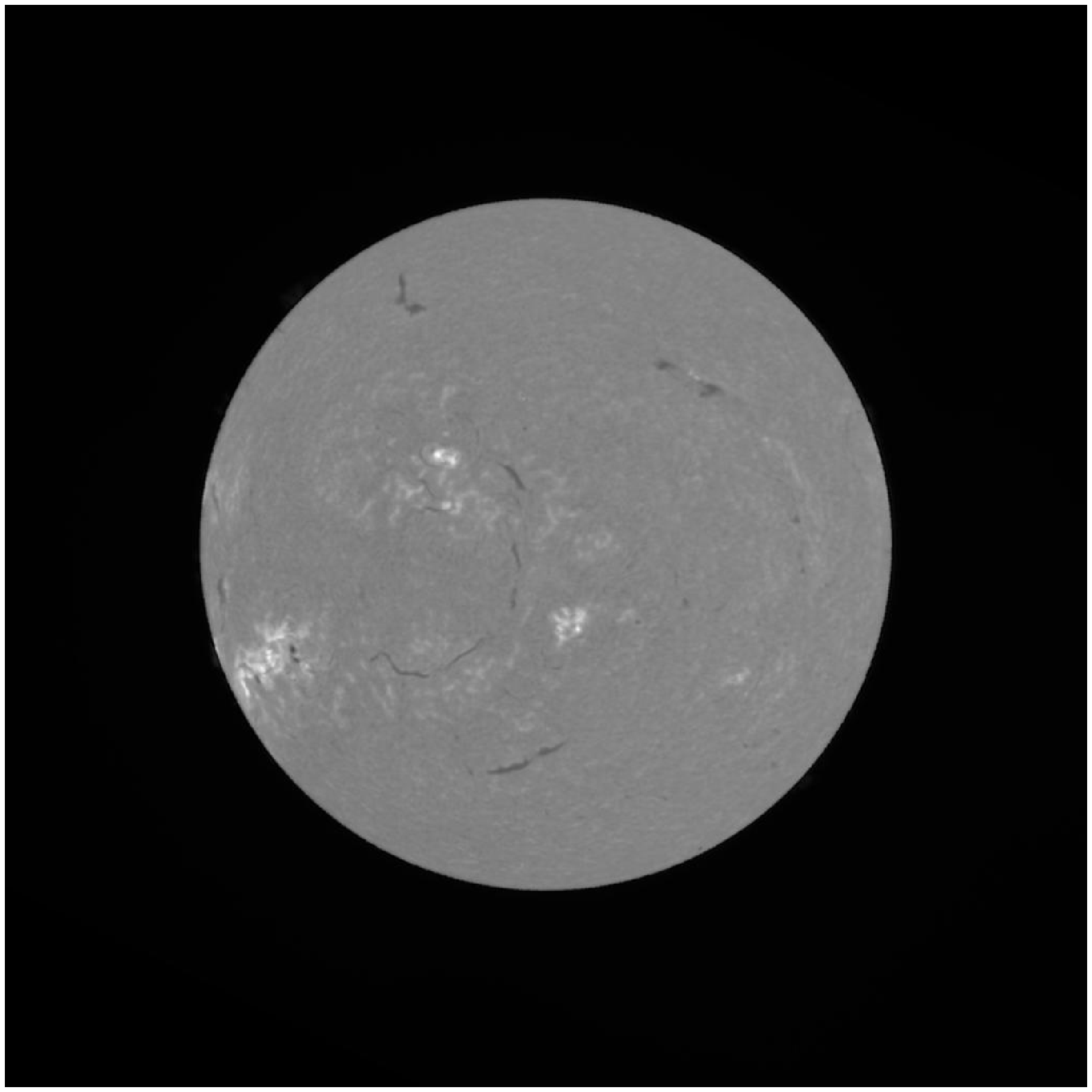}
               \hspace*{-0.075\textwidth}
               \includegraphics[width=0.59\textwidth,clip=]{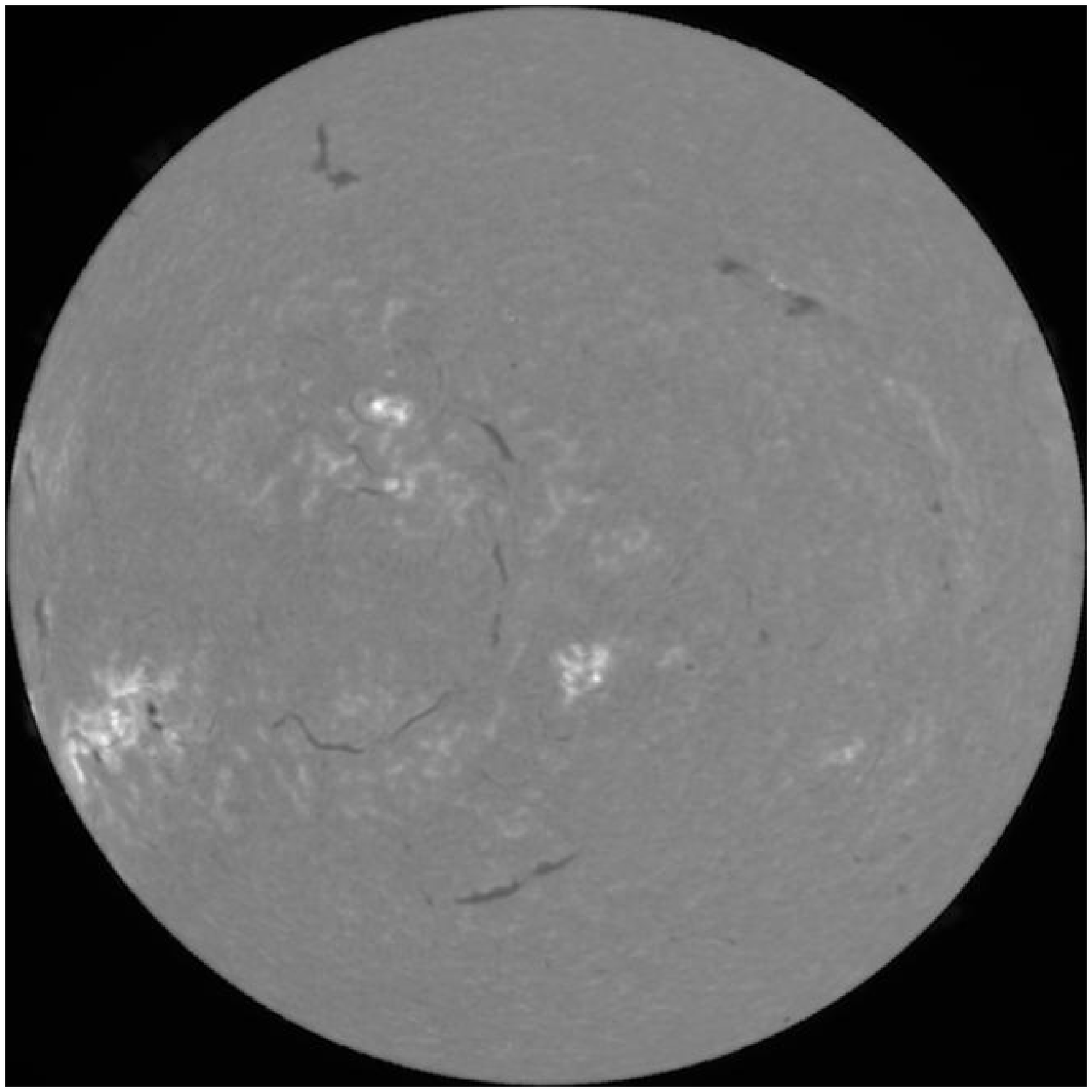}
              }
     \vspace{-0.091\textwidth}
     \centerline{\Large \bf
      \hspace{-0.01 \textwidth} \color{white}{(a)}
      \hspace{0.44\textwidth}  \color{white}{(b)}
         \hfill}
     \vspace{0.02\textwidth}
   \centerline{\hspace*{0.0\textwidth}
               \includegraphics[width=0.59\textwidth,clip=]{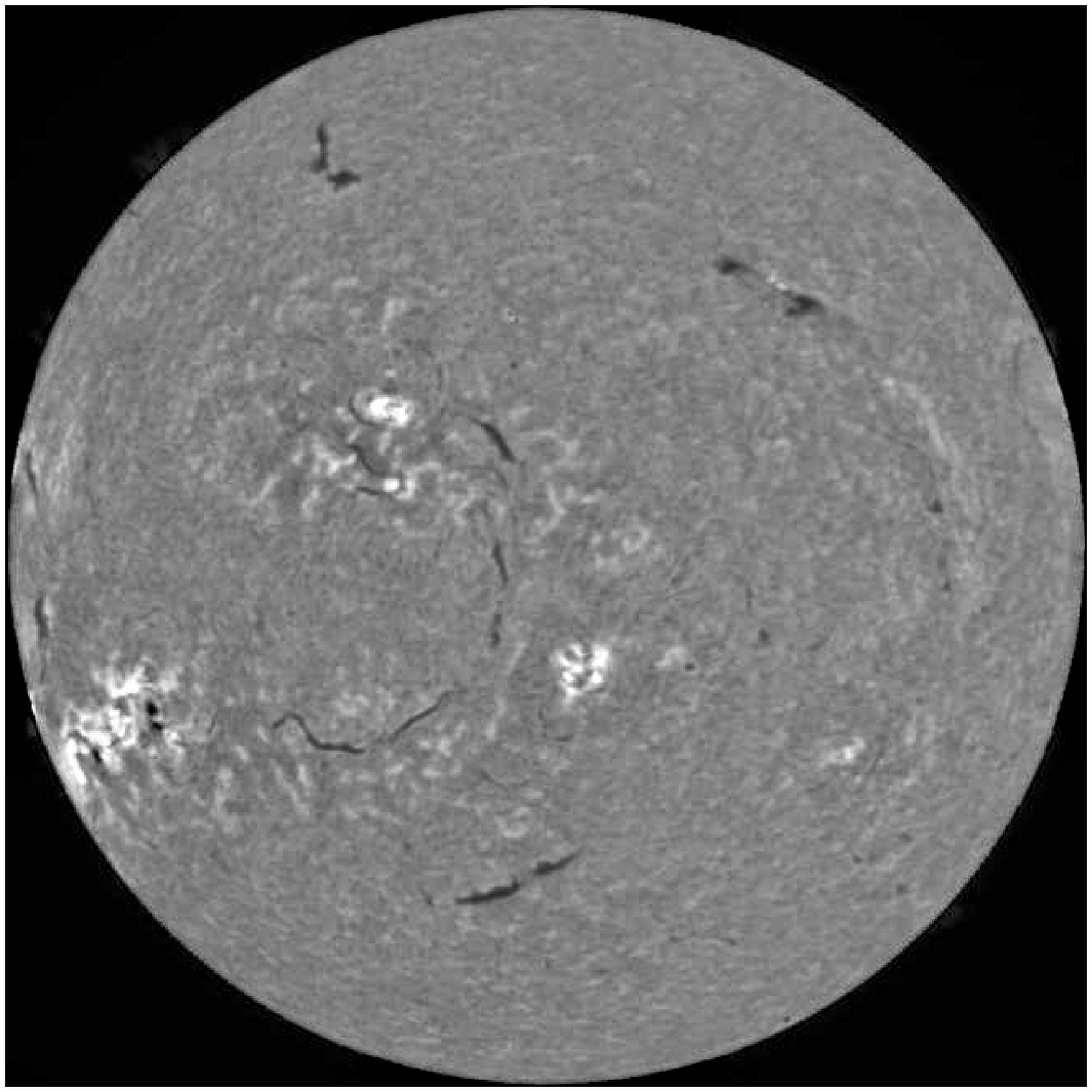}
               \hspace*{-0.075\textwidth}
               \includegraphics[width=0.59\textwidth,clip=]{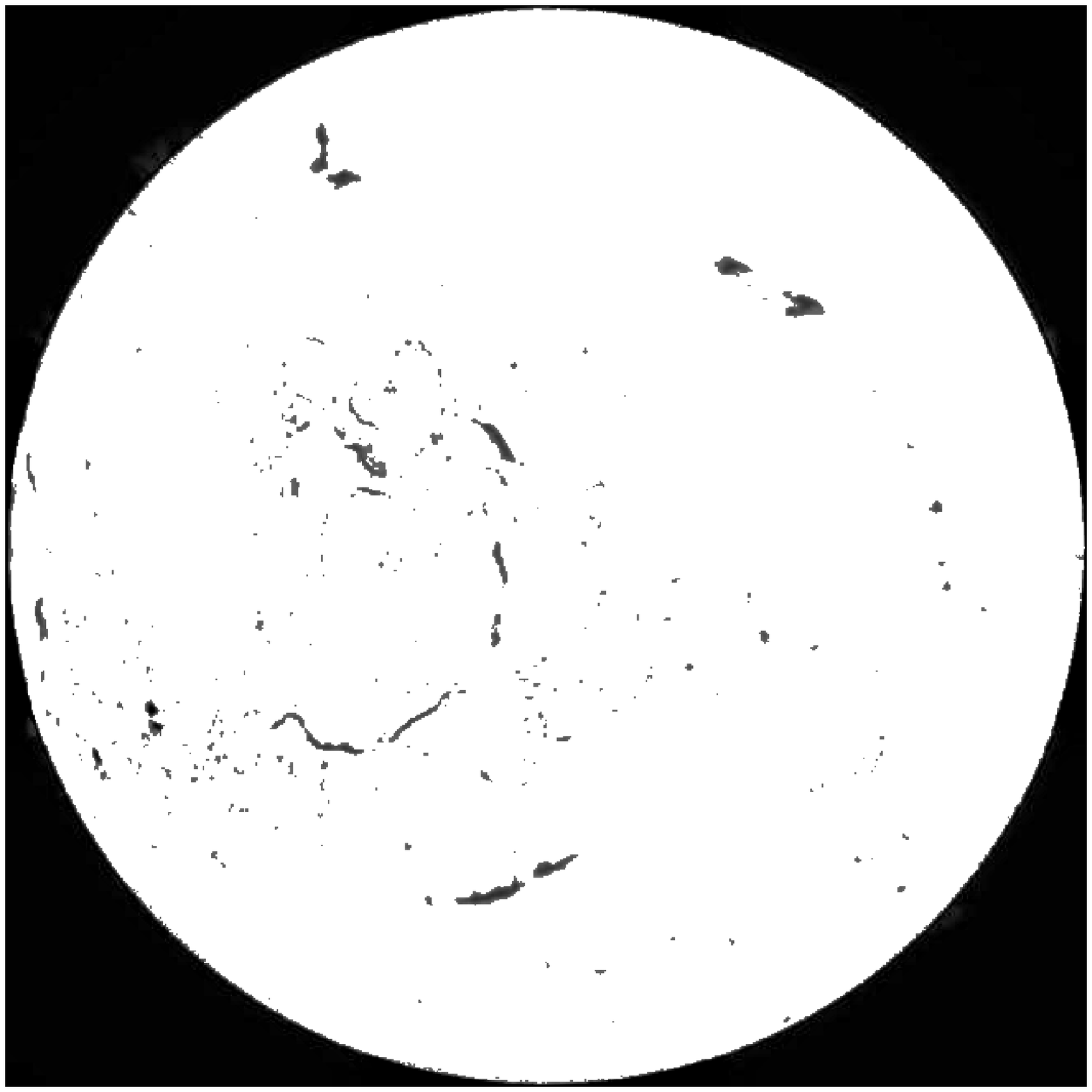}
              }
     \vspace{-0.091\textwidth}
     \centerline{\Large \bf
      \hspace{-0.01 \textwidth} \color{white}{(c)}
      \hspace{0.44\textwidth}  \color{white}{(d)}
         \hfill}
     \vspace{0.02\textwidth}
\caption{An example of the H$\alpha$ image at 18:08:52 UT, 20
April 2001, obtained by Mauna Loa Solar Observatory (MLSO). (a) The
original corrected image downloaded from the MLSO web site. (b) The
sub-image extracted from panel (a). (c) The enhanced image after the top-hat
transformation. (d) The image after a threshold filtering.
        }
   \label{mlso_pre}
   \end{figure}

As a result, we can get an enhanced image, as shown in Figure~\ref{mlso_pre}(c).

\subsection{Threshold Filter}

On the solar disk in H$\alpha$, besides the dark features such as filaments
and sunspots, there are some other bright features such as plages and
flares. Many authors used local threshold method to filter out these
non-interesting features. Actually, after limb-darkening removal and
top-hat filtering,we can easily distinguish the filaments from these
non-filament structures in the gray-scale images via the global
threshold method.We tested several hundred images to find the appropriate
global threshold value. For MLSO images, we found that the threshold value is
about 95--100. The algorithm is simple: if the pixel's value is greater than
the threshold value, we assign it to be 255, which means white in the image. An
example is shown in Figure~\ref{mlso_pre}(d).

\section{Filament Detection}
\label{Detection}

\subsection{Canny Edge Detection}

Segmentation of an image entails the division or separation of the
image into regions of similar attributes \cite{2001Pratt}. The most
basic attribute for segmentation is the intensity level for a gray-scale
image and color components for a color image. In addition,
the image edge is also a useful attribute for segmenting.It is
possible to segment an image into regions of common attribute by
detecting the boundary of each region across which there is a
significant change in intensity. We adopt the most powerful edge-detection
method, \textit{i.e.} the Canny method \cite{Canny},
to identify filaments. The Canny method differs from other edge-detection
methods in that it uses two different thresholds: one for
detecting strong edges and the other for weak edges. The weak edges are
included in the output only if they are connected to strong
edges. Compared to others, this method is therefore less fooled by noise,
and is more likely to detect true weak edges \cite{1990Lim}.
The Canny method works in a multi-step process:

\textbf{Step 1:} Noise reduction. Because the Canny edge detector is
susceptible to noise present in the image data, the image should be smoothed
first. In our method, each image after preprocessing is smoothed by
Gaussian convolution as follows,

   \begin{equation}\label{edge1-equation}
     G(x,y)=\frac{1}{2\pi\sigma^{2}}{\rm e}^{-\frac{x^{2}+y^{2}}{2\sigma^{2}}}
     \, ,
   \end{equation}

   \begin{equation}\label{edge2-equation}
     f'(x,y)=\sum_{x'}\sum_{y'} f(x,y) G(x-x',y-y') \, ,
   \end{equation}

\noindent
where $G(x, y)$ is the 2D Gaussian filter; $f(x, y)$ is the
input image and $f'(x, y)$ is the output image which is convolved
with the 2D Gaussian filter; ($x, y$) is the position in the $x$--$y$ plane
of the image. We choose $\sigma = 1$ in our processing.

\textbf{Step 2:} Finding gradients. The edges usually can be found at
those places where the gray-scale intensity drastically
changes. It means that we can find them by checking the gradient at each
pixel in the image. The first is to get the gradient in the $x$-direction
$g_{x}(x, y)$ and $y$-direction $g_{y}(x, y)$, respectively, by
applying the derivative of a Gaussian filter:

   \begin{equation}\label{edge3-equation}
     g_{x}(x,y)=\sum_{x'}\sum_{y'} f(x,y) g_{1}(x-x',y-y') \, ,
   \end{equation}
   \begin{equation}\label{edge4-equation}
     g_{y}(x,y)=\sum_{x'}\sum_{y'} f(x,y) g_{2}(x-x',y-y') \, ,
   \end{equation}

\noindent
where
 \begin{equation}\label{edge5-equation}
  g_{1}(x,y)=-\frac{1}{\pi\sigma^{2}}x{\rm e}^{-\frac{x^{2}+y^{2}}{2\sigma^{2}}}
  \, ,
 \end{equation}

 \begin{equation}\label{edge6-equation}
   g_{2}(x,y)=-\frac{1}{\pi\sigma^{2}}y{\rm e}^{-\frac{x^{2}+y^{2}}{2\sigma^{2}}} .
 \end{equation}

Then we use the following two equations to determine the gradient magnitude
and the direction of the edge:

   \begin{equation}\label{gradient-equation}
     g(x,y)=\sqrt{g_{x}(x,y)^{2}+g_{y}(x,y)^{2}} \, ,
   \end{equation}

   \begin{equation}\label{gradient-dir-equation}
     \theta(x,y)=\arctan\frac{g_{y}(x,y)}{g_{x}(x,y)} \, .
   \end{equation}

\textbf{Step 3:} Non-maximum suppression. For an image array, the
edge direction angle is rounded to one of four angles representing vertical,
horizontal and the two diagonals (\textit{i.e.} 0, 45, 90, 135, 180, 225, 270, 315 and 360 degrees),
corresponding to the use of an 8-connected neighbourhood. Then, for
each pixel of the gradient image, we compare the edge gradient
magnitude of the current pixel with the edge gradient magnitude of
the pixel along the gradient direction. For example, if the gradient
direction is to the northeast, the pixel should be compared with the pixels
to the northeast and to the southwest. If the edge gradient magnitude of
the current pixel is the largest one, we mark it as one part of the edge.
If not, we suppress it, \textit{i.e.} it is ignored.

\textbf{Step 4:} Edge tracing by hysteresis. After step 3, many of
the remaining edge pixels would probably be the true edges of filaments,
but some may be caused by noises. The Canny method uses thresholding
with hysteresis to determine whether the edges obtained in step 3 are true or
not. The algorithm adopts two thresholds, \textit{i.e.} high and low
thresholds: If the edge pixel's gradient magnitude is higher
than the high threshold, the pixel is marked as a strong pixel; if the edge
pixel's gradient magnitude is lower than the low threshold, the
pixel will be suppressed; if it is between the two thresholds, the
pixel will be marked as a weak one.

In order to find the thresholds, we use an automatic method: First, we should
provide a probability of the pixels that are not the edge points and calculate
the number of pixels that may not be the edge points in the entire image by the
probability; Then we increase the gradient threshold until the total number of
the pixels with the gradient smaller then the threshold is just greater than
the probability value, then the current gradient threshold is chosen as the
high threshold. The low threshold is about half of the high one. In our process
the probability is chosen to be about 0.98, and the low threshold is 0.4 times
the high threshold.

After tracing through the entire image we have strong and weak pixel arrays
which can be treated as a set of edge curves. The weak edges are included if
and only if they are connected to strong edges. We scan the entire binary
image to find the pixels where strong and weak edges overlap each other and
finally get the edge map. Then, the morphological thinning operation
\cite{1992Lam} is applied to minimize the connected lines in order to get
accurate and fine edges.

\begin{figure}    
   \centerline{\hspace*{0.0\textwidth}
               \includegraphics[width=0.59\textwidth,clip=]{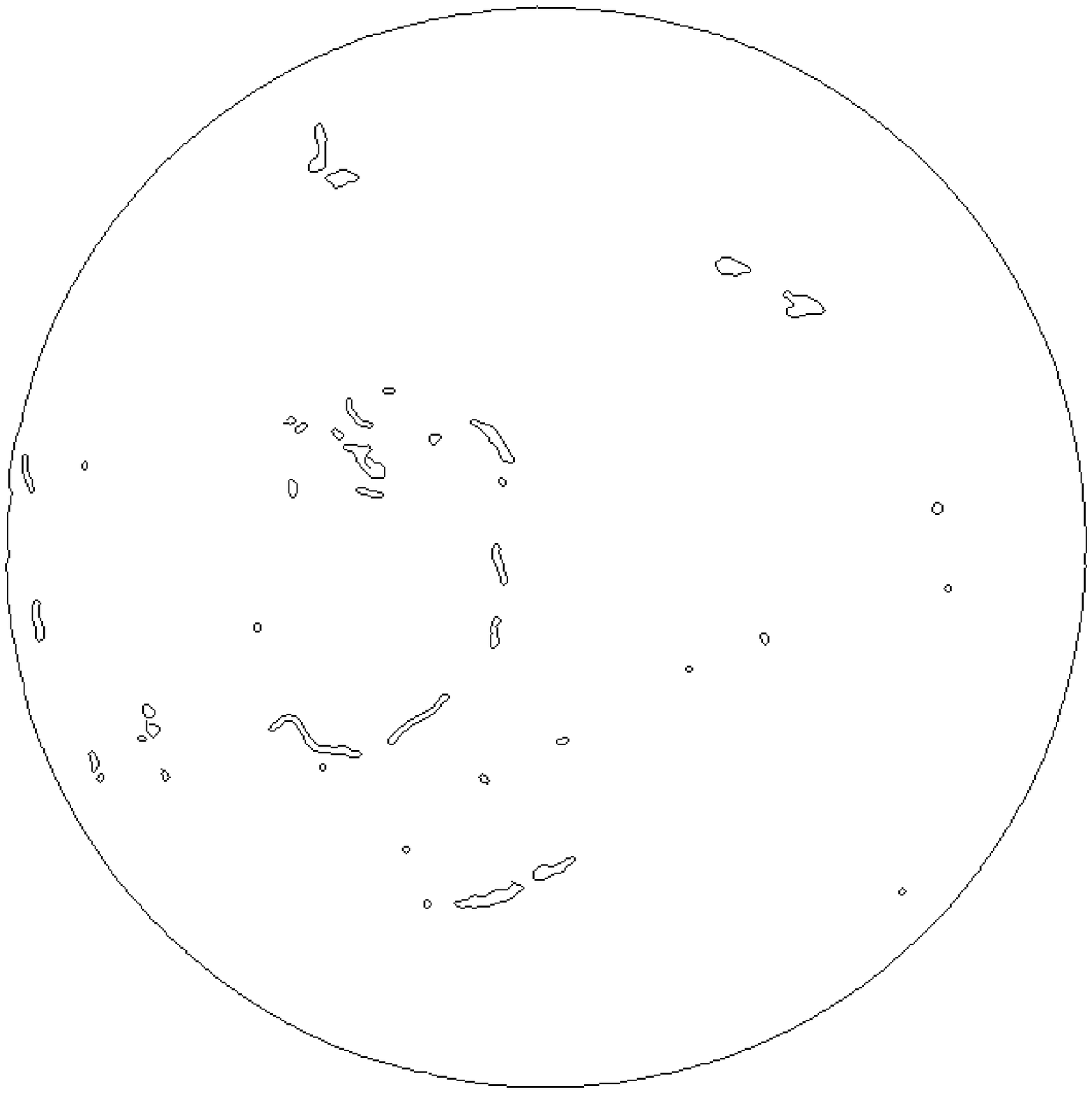}
               \hspace*{-0.075\textwidth}
               \includegraphics[width=0.59\textwidth,clip=]{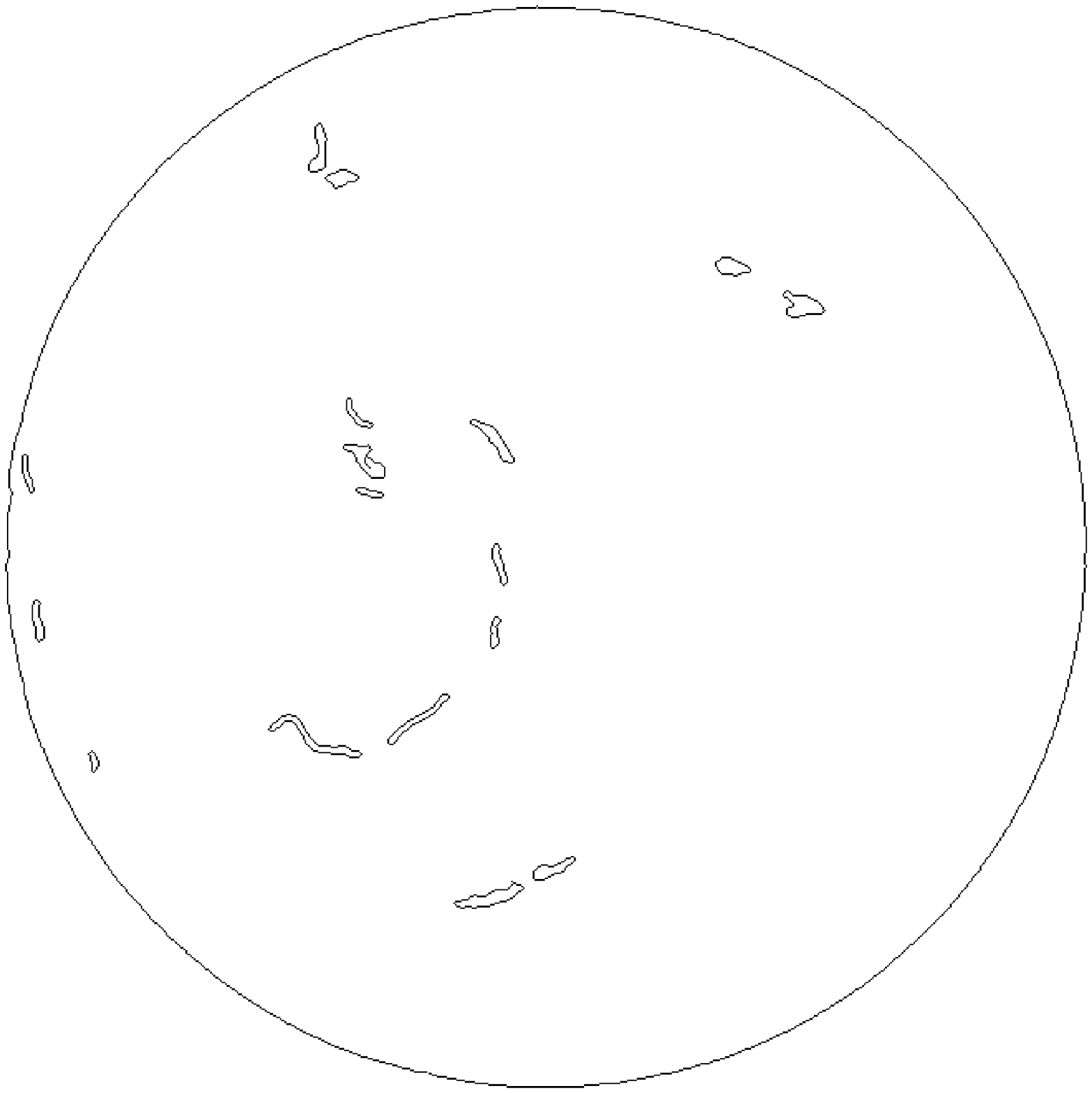}
              }
     \vspace{-0.091\textwidth}
     \centerline{\Large \bf
      \hspace{-0.01 \textwidth} \color{black}{(a)}
      \hspace{0.44\textwidth}  \color{black}{(b)}
         \hfill}
     \vspace{0.02\textwidth}
\caption{Examples after filament detection processing. (a) After
Canny edge detection, the filaments were segmented, the result shows
the edge of each filament or fragment.  (b) The image after sunspot
removal. Compared with (a), small cycles (\textit{i.e.} the edges of sunspots) were removed. In order to obtain clear results, the foreground and background colors are exchanged, \textit{i.e.} the filament features become black and the background becomes white. The same is for Figure \ref{mlso_feature}.
        }
   \label{mlso_detection}
   \end{figure}

After applying the Canny edge-detection method, we get the edge of each
filament, as shown in Figure~\ref{mlso_detection}(a).

\subsection{Connected Component Process}

\begin{figure}    
   \centerline{\hspace*{-0.0\textwidth}
               \includegraphics[width=0.8\textwidth,clip=]{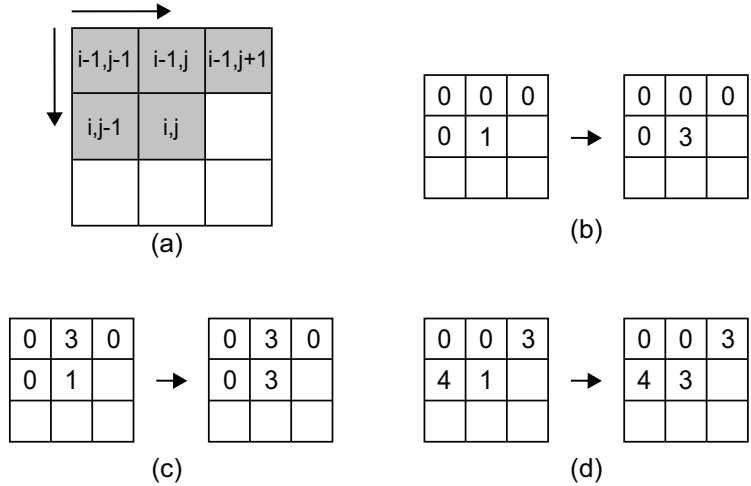}
               \hspace*{0.0\textwidth}
              }
     \vspace{0.01\textwidth}

\caption{Scanning sequence (panel a) and three examples for the
connected components label method (panels b, c, and d).
        }
   \label{connect_component_label}
   \end{figure}

After segmentation process, in the computer vision the image is still an array
of pixels. We are not interested in each pixel but the special region
(\textit{i.e.} the filament) constituted by the pixels. These regions are
called the connected components of the binary images, which are more
complex and have a rich set of properties (\textit{i.e.} shape, position,
and area). We use a classical method \cite{1992Haralick} to realize the
connected component labeling. It means that the pixels in a connected component
are given with the same identity label. After connected component labeling, the
product we get is changed from pixels to regions that we are interested in. The
input binary images are of 8-connectivity, and the algorithm consists of the
following two steps:

In the first step, the algorithm goes through each pixel from left to right
and from top to bottom, as indicated by the arrows shown in Figure~\ref{connect_component_label}(a).
It checks the labels of four neighboring pixels that are north-east, north,
north-west and west of the current pixel. For example, suppose the current pixel is $(i,j)$ as shown in Figure~\ref{connect_component_label}(a), the code checks the labels of four pixels that are $(i-1,j+1)$,
$(i-1,j)$, $(i-1,j-1)$, and $(i,j-1)$ :

i) If all four neighbors are not assigned, a new label is assigned to
the current pixel. An example is shown in Figure~\ref{connect_component_label}(b): Supposing
that the value of the current pixel is 1 and the values of the
four neighbors are 0 (0 means this pixel is a background pixel, and 1 means the pixel is the foreground), it means a new filament is encountered. If the
previous label is ``2", we assign label ``3" to the new filament.

ii) If one of the four neighbors has been labeled before, we assign
  the neighbor's label to the current pixel. An example is shown in Figure~\ref{connect_component_label}(c):
One of the four neighbors has been labeled, \textit{i.e.} the north neighbor has been labeled ``3", so we assign the same label to the current pixel.

iii) If more than one of the neighbors have been labeled before, we
assign the smaller label to it. An example is shown in
Figure~\ref{connect_component_label}(d): Two neighbors have been labeled,
\textit{i.e.} the north-east and west neighbors. The label of north-east
neighbor is ``3", which is smaller than the west neighbor's, so we use "3"
to assign the current pixel.

After completing the scanning, the equivalent label pairs are sorted
into equivalence classes and a unique label is assigned to each
class.

In the second step, the above algorithm goes through again,
during which each label is replaced by the label assigned to its
equivalence class. After completing the scanning, a unique label is
assigned to each equivalence class. In other words, we have assigned
each filament a unique label.

\subsection{Sunspot Filter}

Actually, it is not easy to distinguish between sunspots and filaments by
gray-scale levels. The labeled ``filaments" so far may include some sunspots.
Thus, we have to separate real filaments from sunspots by use of geometric
structures: the size and the long-to-short-axis ratio of the filament.A
candidate with the size (e.g. the perimeter) larger than the threshold is
considered to be a filament. In the case that the size of the labeled object is
smaller than the size threshold, only if the ratio of the long axis to short
axis is larger than a given value, the candidate is treated as a small
filament; otherwise it would be removed. We have got the edge of each filament
whose length can be treated as the filament perimeter. We set the perimeter
threshold to be 25 pixels, and the long-to-short-axis ratio threshold being 2
in our procedure. An example is shown in Figure~\ref{mlso_detection}(b). The
filament label should be updated after removing the sunspots. Then each
filament is labeled with a unique number. An example is shown later in
Figure~\ref{mlso_update}(a). This method is used to filter out sunspots. With
the same method, we can filter out other features. In other words, we can adopt
the method to automatically detect sunspots, which will be implemented in our
future work.

\section{Filament Feature Recognition}
\label{Feature}

\subsection{Perimeter}

As mentioned above, we have the filament edge, which can be easily used to
derive the filament perimeter after the connected component process.
It is done by the integration of the distance connecting neighboring pixels
along the edge of each filament.

\subsection{Position}

We choose the geometric center of each filament, \textit{i.e.} the centroid,
as the location of the filament. First, we find the centroid of a filament
($x_{c}, y_{c}$), {\it i.e.} we calculate the average of the abscissa and
the ordinate of all the filament pixels. Since filaments follow the solar
rotation and the rotation axis wobbles with time, the position of a certain
filament has an elliptic orbit in the plane of the sky. In order to get the
longitude and the latitude of the filament in the heliographic coordinates, we
use the Solar SoftWare routine ``xy2lonlat.pro".

It is noted that the center of the solar disk we used here is not that
provided in the header of the ``FITS" file. After the Canny edge detection,
besides the locations of the filaments, we also got the boundary of the
solar disk, and it is the biggest connected component after the
connected component processing. We fit the circle and calculate the
geometric center of the fitted circle, which is the exact center of
the solar disk.

\begin{figure}    
   \centerline{\hspace*{0.0\textwidth}
               \includegraphics[width=0.59\textwidth,clip=]{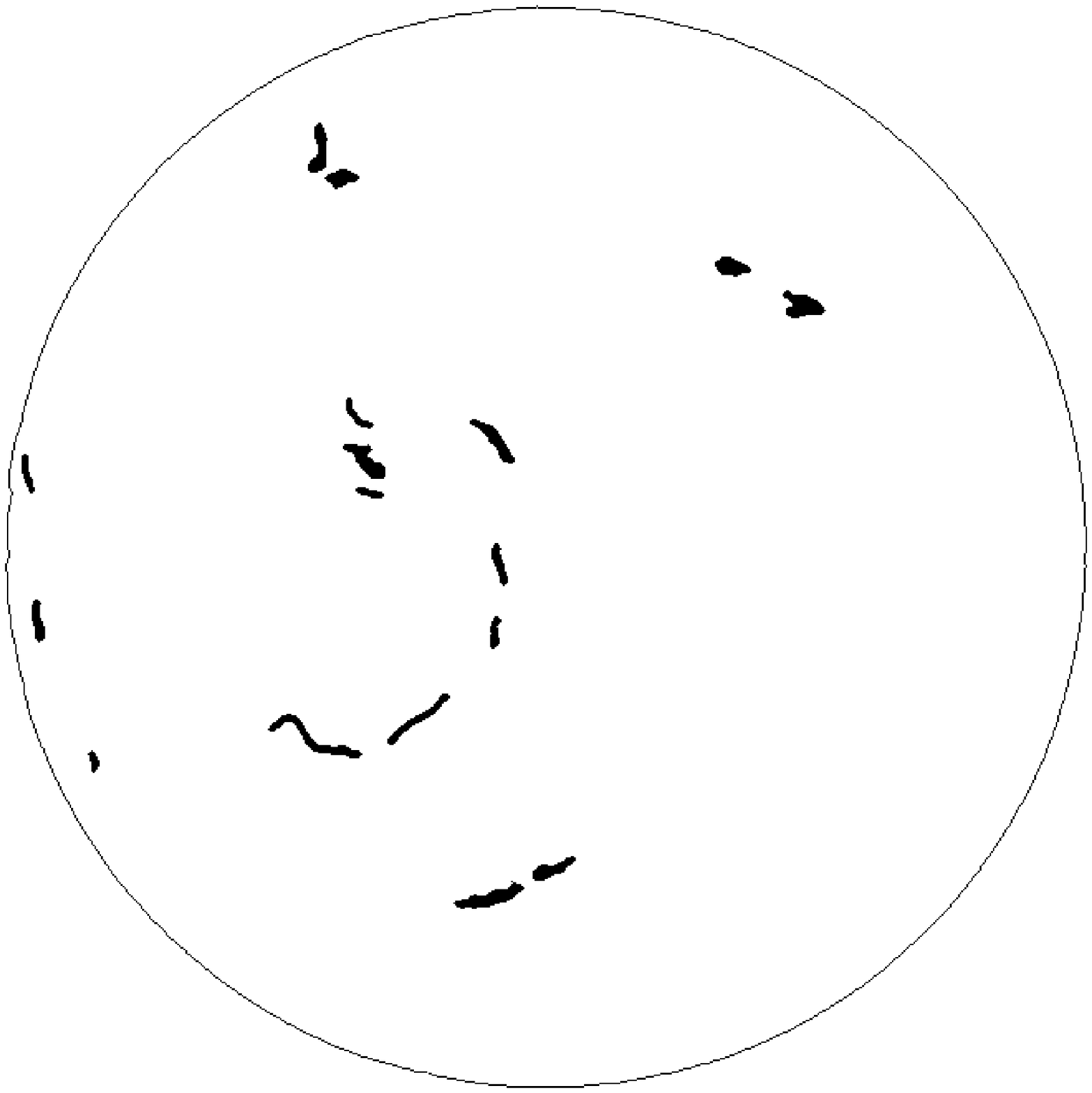}
               \hspace*{-0.075\textwidth}
               \includegraphics[width=0.59\textwidth,clip=]{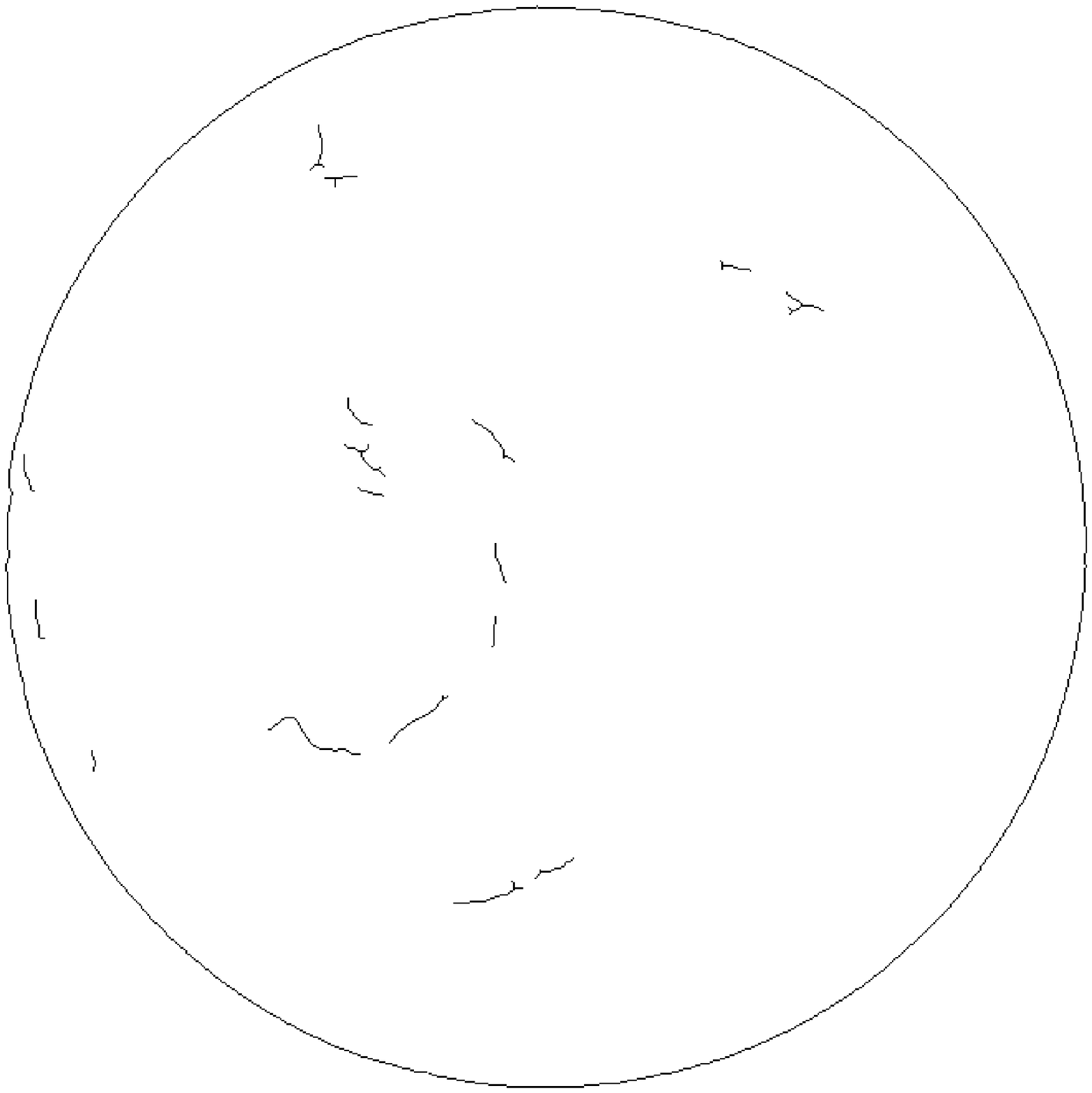}
              }
     \vspace{-0.091\textwidth}
     \centerline{\Large \bf
      \hspace{-0.01 \textwidth} \color{black}{(a)}
      \hspace{0.44\textwidth}  \color{black}{(b)}
         \hfill}
     \vspace{0.02\textwidth}
\caption{Examples of filament feature recognition. (a) The image after
morphological reconstruction processing,  the edge curve is full
filled with white pixels, from which we get the filament area.
(b) The filament spines after morphological skeletonization and
barbs removal processing.
        }
   \label{mlso_feature}
   \end{figure}

\subsection{Area}

In our process, the area of a filament is the integration of the pixel area
divided by the cosine of the heliocentric angle, where the integration is taken
in the area enclosed by the edge curve. We use the foreground color (white) to
fill the edge curve so that the pixels inside the curve become white, then the
white pixels constitute the filament area. The algorithm is based on
morphological reconstruction \cite{Soille_1999}. The edge curve with holes is
filled with white pixels. The hole is a set of background pixels (\textit{i.e.}
black pixels) surrounded by foreground pixels (white pixels). An example is
given in Figure~\ref{mlso_feature}(a) and an extracted filament example is
shown in Figure~\ref{feature_example}(c).

\subsection{Spine}

\begin{figure}    
   \centerline{\hspace*{0.0\textwidth}
               \includegraphics[width=0.8\textwidth,clip=]{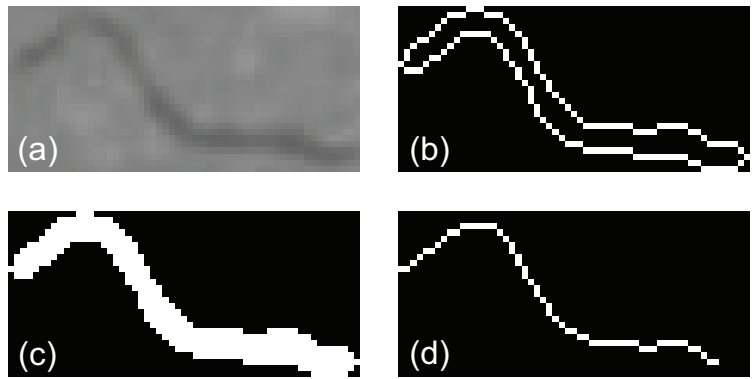}
              }

     \vspace{0.02\textwidth}
\caption{An example of a filament feature recognition. (a) A filament
extracted from the original image (\textit{i.e.} Figure~\ref{mlso_pre}(a)).
(b) The filament edge (\textit{i.e.} the perimeter). (c) The filament edge was filled with white pixels, from which we got the filament area. (d) The filament skeleton (\textit{i.e.} spines).
        }
   \label{feature_example}
   \end{figure}

The morphological skeletonization method is adopted to get the filament spine,
which is a stick-like skeleton, spatially placed along the medial region of the
filament. The skeleton is a unit-wide set that contains only the pixels that
can be removed without changing its topology \cite{Kong_1996}. Iterations of
the morphological-thinning operation are employed for the skeleton method
\cite{1992Haralick}. Assuming that the input binary image is $I$ (\textit{i.e.}
a matrix only has values of 0 and 1), the details of the method are described
below:

First, the thinning processing of $I$ by structuring-element pair $(J, K)$ is defined as

   \begin{equation}
   I\oslash(J, K) = I - I\otimes(J, K) \, ,
   \end{equation}

\noindent
where $\oslash$ is the morphological thinning operator, $J$ and $K$ must
satisfy $ J \cap K = \emptyset$, where the symbol $\emptyset$ represents the
empty set. $I\otimes(J, K)$, \textit{i.e.} the hit-and-miss transformation of
set $I$ by $(J, K)$, is defined by

   \begin{equation}
   I\otimes(J, K) = (I \ominus J)\cap(I^{c} \ominus K) \, ,
   \end{equation}

\noindent
where $\otimes$ is the hit-and-miss transformation operator, $\cap$ the
intersection operator, $I^{c}$ the complement of $I$ and the symbol
$\ominus$ denotes the morphological erosion operator.

\begin{figure}
\centerline{\includegraphics[width=0.7\textwidth,clip=]{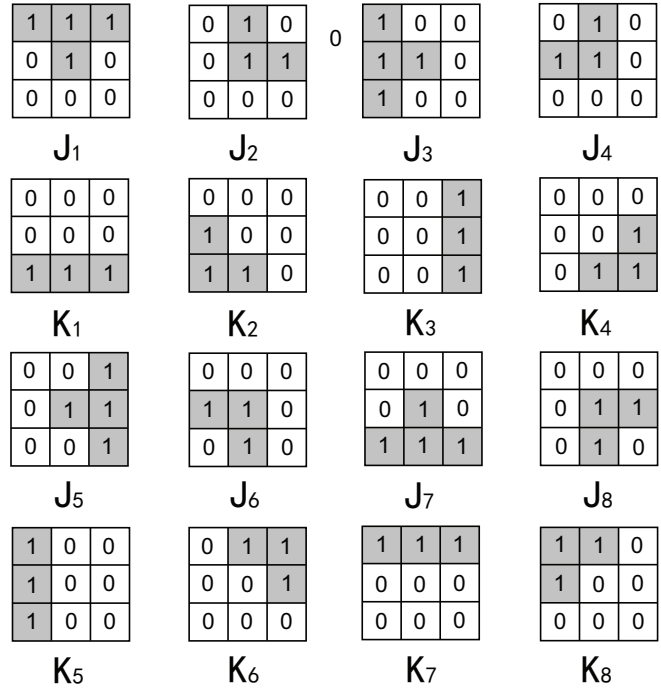}}
\caption{Structuring-element pair $(J_{i}, K_{i})$ used to determine the
filament skeleton. $J_{i}$ and $K_{i}$ must satisfy $ J_{i} \cap K_{i} = \emptyset$.
}
\label{jk_array}
\end{figure}

Then, using the sequence of eight structuring-element pairs as shown
in Figure~\ref{jk_array} to iteratively process the thinning operation:

   \begin{equation}
    I_{N+1} = ( \ldots \{ [I_{N}\oslash (J_{1},K_{1})] \oslash (J_{2},K_{2})\}
    \oslash \ldots \oslash (J_{8},K_{8}) ) .
   \end{equation}
We let $I_{0} = I$, which is firstly thinned by the structure-element pair
$(J_{1}, K_{1})$, and then by $(J_{2}, K_{2})$, \ldots, $(J_{8}, K_{8})$, the
thinned result is defined as $I_{1}$. Such a process is repeated in order to
get $I_{2}$, $I_{3}$, \ldots, $I_{N}$.

Finally, the thinning process repeats until $I_{N}=I_{N+1}$, \textit{i.e.} the
filament skeleton is the final structure that can not be thinned any more.
Actually, some of the resulting filament skeletons still contain small barbs.
We use the morphological hit-and-miss transformation to find the endpoints that
constitute the barbs, and then remove the barbs.
This process may iterate
several times because some of these points may not be removed in one go. In our
process it is iterated four times. The filament spines on the whole Sun are
shown in Figure~\ref{mlso_feature}(b) and an extracted filament example is
shown in Figure~\ref{feature_example}(c).

\subsection{Tilt Angle}

The tilt angle is defined as the fitted filament spine orientation
with respect to the solar Equator. After we get the spine, we use
a linear polynomial to fit it, by which we calculate the slope $A$.
The tilt angle is $\arctan(A)$.

\subsection{Feature Update}

\begin{figure}    
   \centerline{\hspace*{0.0\textwidth}
               \includegraphics[width=0.59\textwidth,clip=]{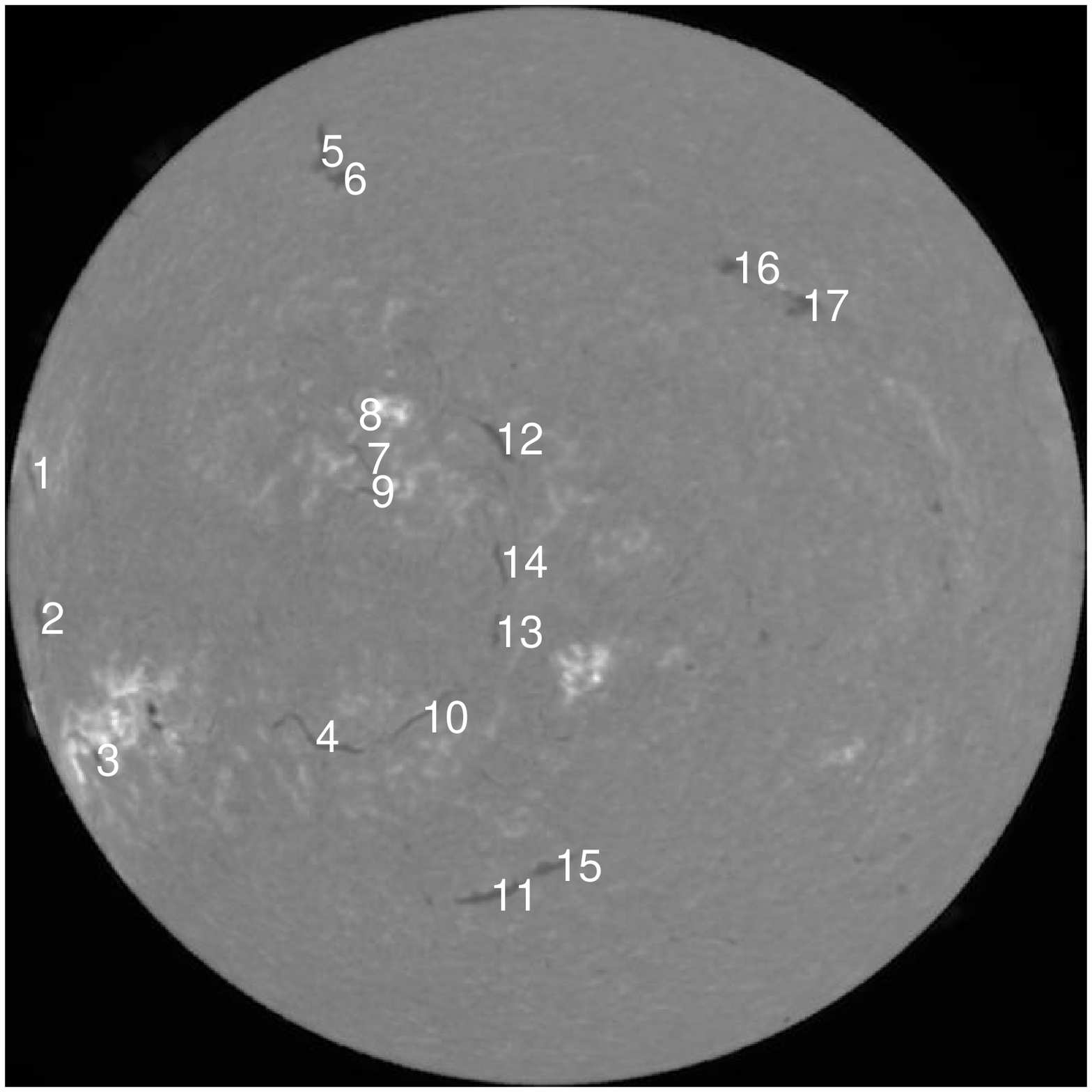}
               \hspace*{-0.075\textwidth}
               \includegraphics[width=0.59\textwidth,clip=]{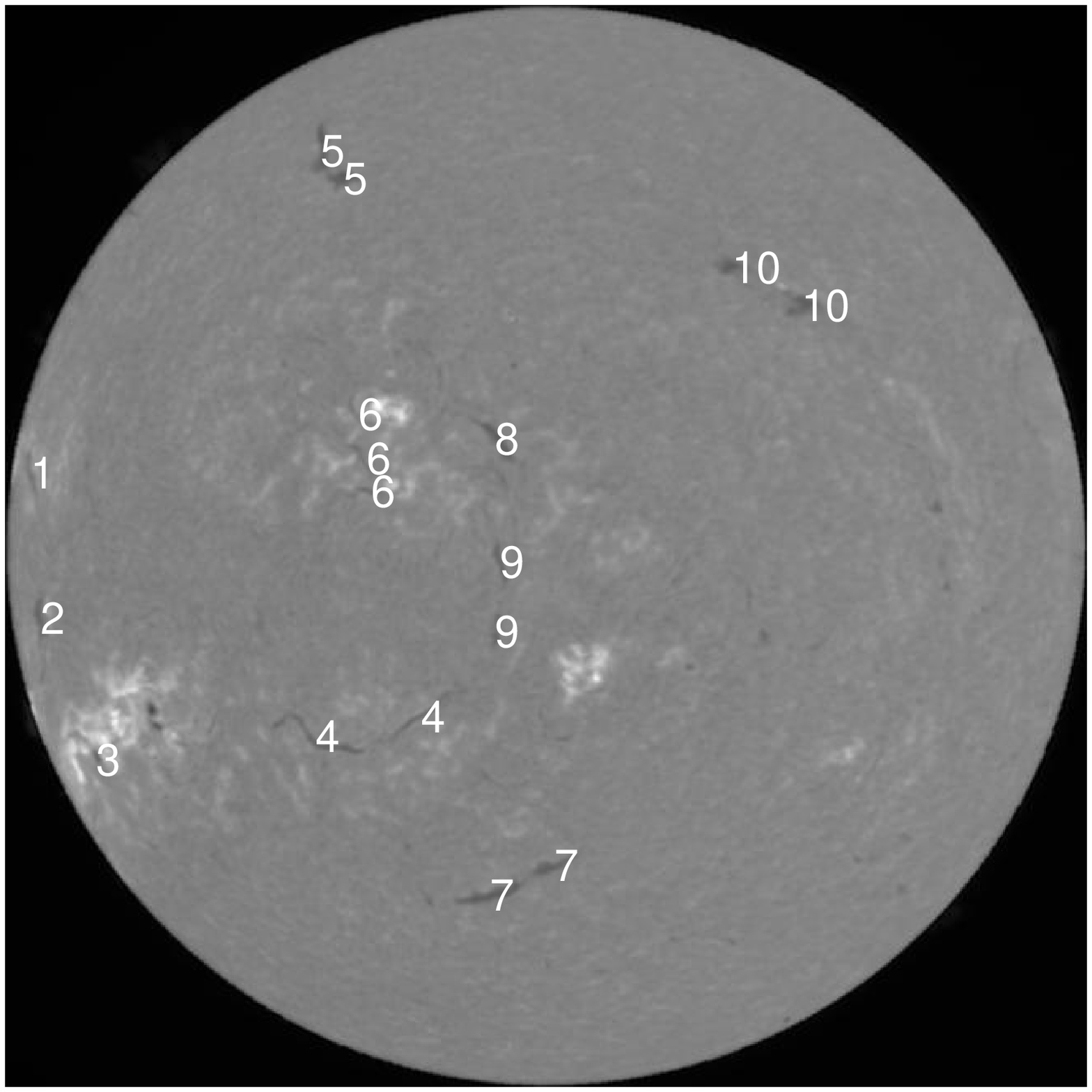}
              }
     \vspace{-0.091\textwidth}
     \centerline{\Large \bf
      \hspace{-0.01 \textwidth} \color{white}{(a)}
      \hspace{0.44\textwidth}  \color{white}{(b)}
         \hfill}
     \vspace{0.02\textwidth}
\caption{An example of the detection results. (a) After filament detection,
each filament or fragment is labeled with a unique number. It shows there are in total 17 filaments or fragments, and each of their centers is labeled with a unique number. (b) Each candidate filament that we obtained after the filament
detection processing may not be a true one but a filament fragment. After the
feature update processing, the fragments belonging to a single filament are
labeled with the same number. For example, filament fragments numbers 11 and 15
in (a) were recognized as one filament, thus they are updated with the same label number 7 in panel (b).
}
   \label{mlso_update}
   \end{figure}

A filament may consist of several fragments. Thus, the filaments we obtained
after the previous processing may not be the real individual filaments, with
some being fragments of one filament. We adopt a ``distance criterion" in order
to find the fragments belonging to a single filament. The
method is the same as the ``labelling criterion" filament tracking method used
by \inlinecite{Joshi2010}, which is explained as follows: For a certain
filament or filament fragment, we compare it with all other
fragments. The fragments would be recognized to belong to a common
filament if the two fragments lie within the distance threshold. The process
is iterated until all fragments are checked.
The filaments in the new image are compared with those in the previous image. The experiential distance threshold in our processing is taken to be
60 pixels for the MLSO data. After this process, the fragment labels will be
updated; if several fragments belong to a common filament, the label
should be unified, as shown in
Figure~\ref{mlso_update}(b). For example, filament fragments number 11
and 15 in Figure~\ref{mlso_update}(a) are recognized as one
filament, thus they are updated with the same label number 7. For the update of
the area, perimeter, and the length of spine, we just calculate the
sum of each fragment with the same label, while the position and the
tilt angle must be reprocessed.

Another criterion for identifying a broken filament is to compare the tilt
angle of the fragments, \textit{i.e.} if the neighboring candidates have
similar tilt angles, they could be considered as one big filament. This method
works fine for the magnetic inversion lines that are not strongly curved, and
will be incorporated in our future version.

\begin{figure}    
   \centerline{\hspace*{0.0\textwidth}
               \includegraphics[width=0.59\textwidth,clip=]{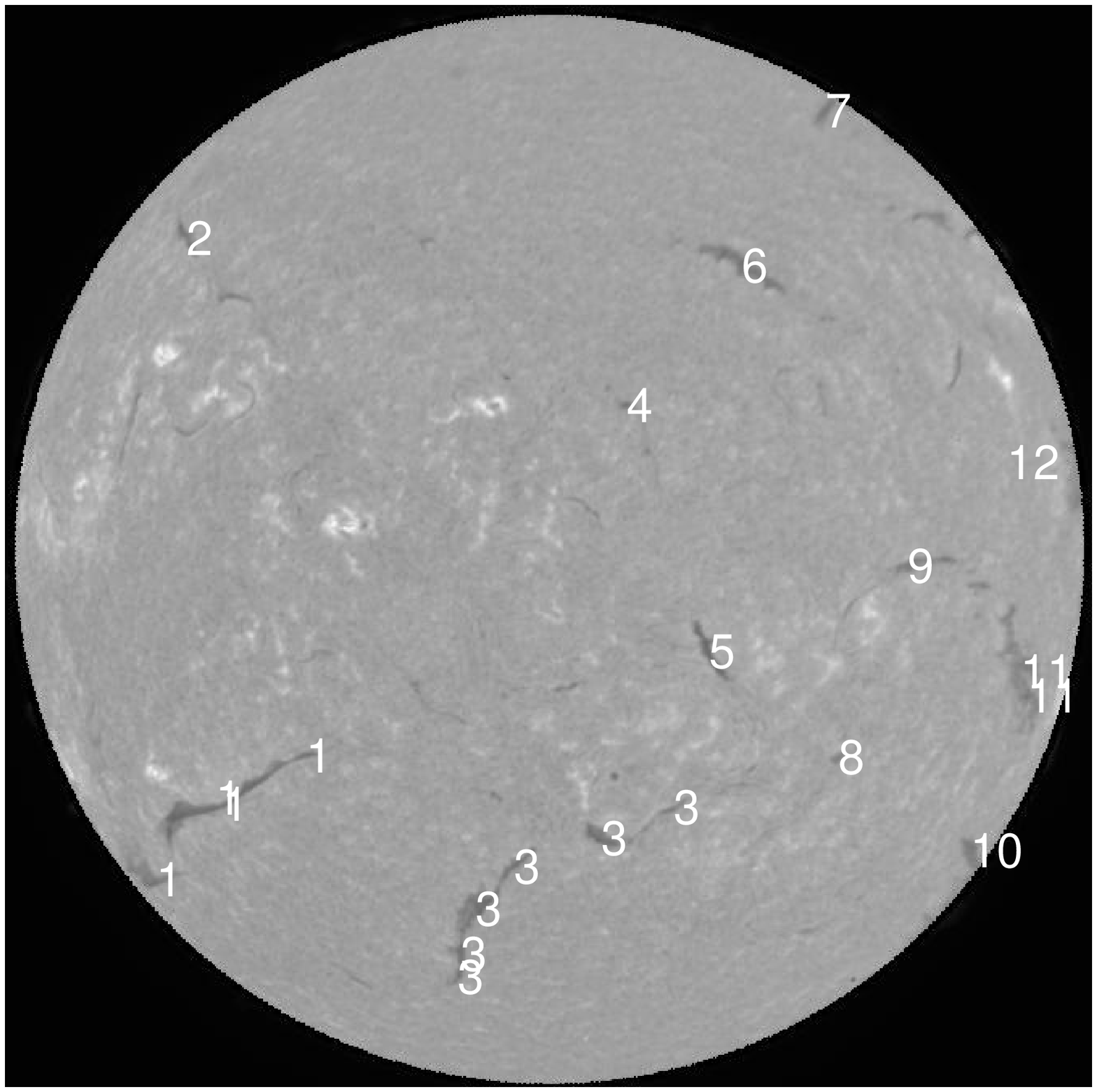}
               \hspace*{-0.075\textwidth}
               \includegraphics[width=0.59\textwidth,clip=]{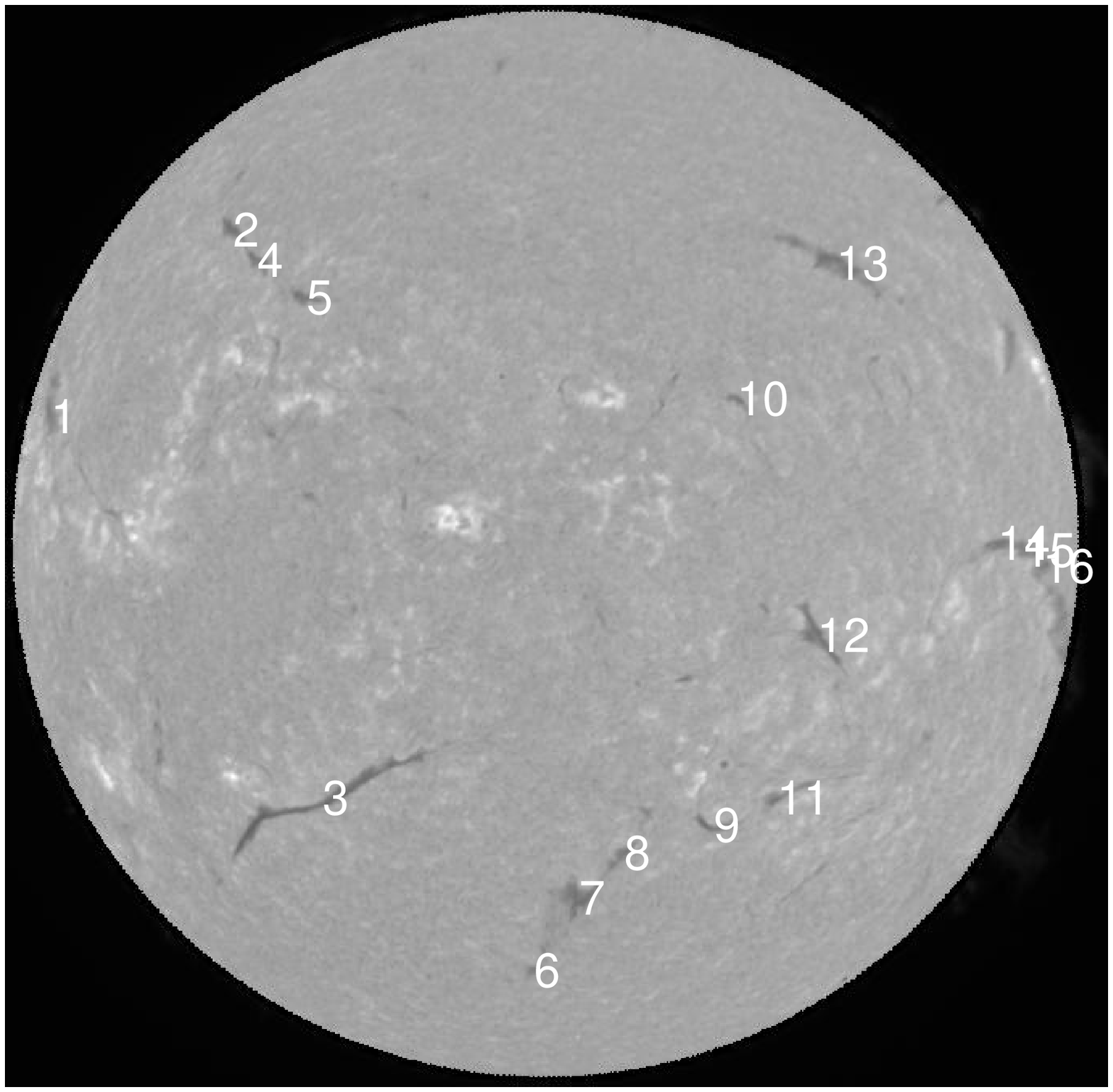}
              }
     \vspace{-0.091\textwidth}
     \centerline{\Large \bf
      \hspace{-0.01 \textwidth} \color{white}{(a)}
      \hspace{0.44\textwidth}  \color{white}{(b)}
         \hfill}
     \vspace{0.02\textwidth}
   \centerline{\hspace*{0.0\textwidth}
               \includegraphics[width=0.59\textwidth,clip=]{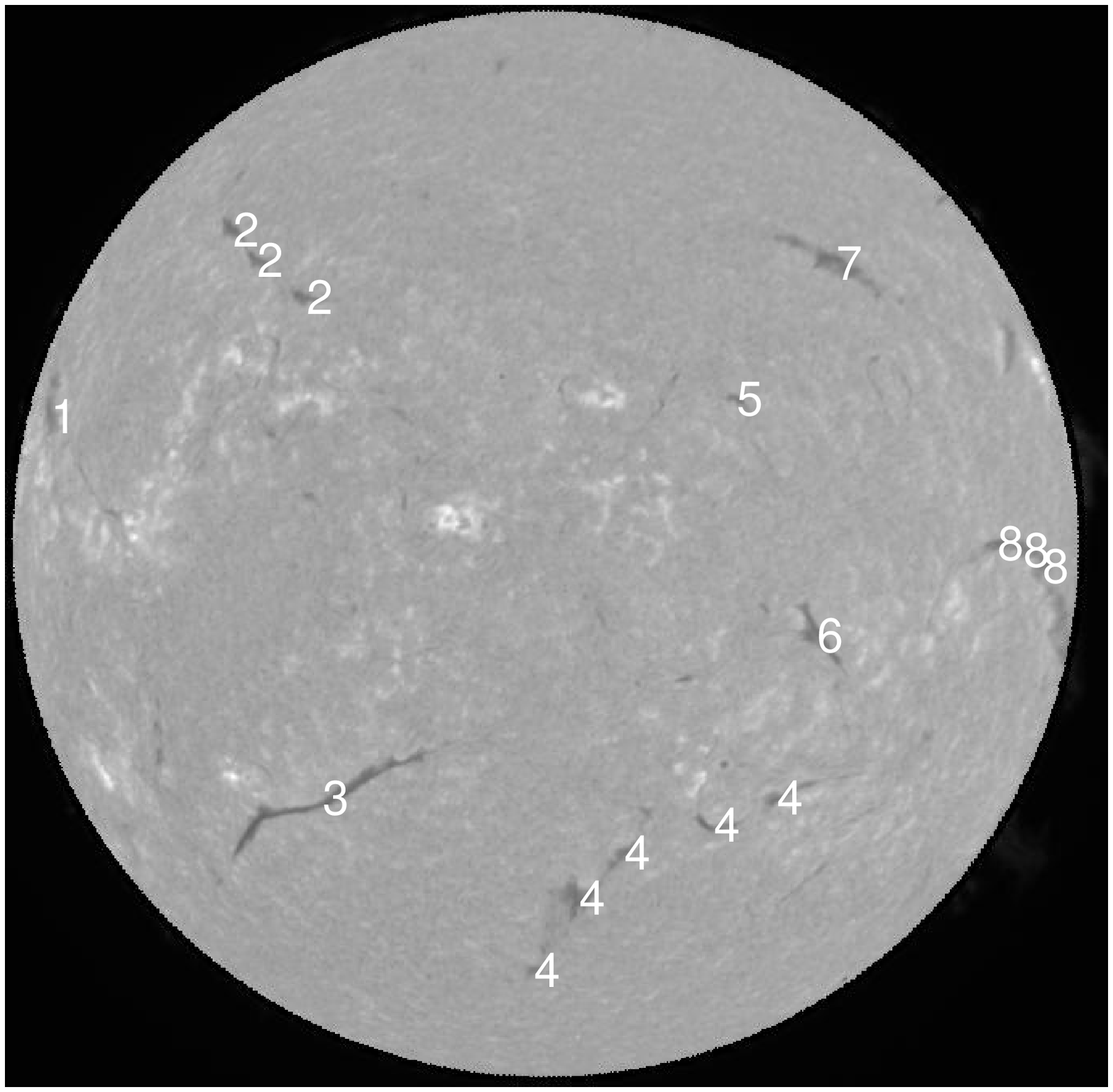}
               \hspace*{-0.075\textwidth}
               \includegraphics[width=0.59\textwidth,clip=]{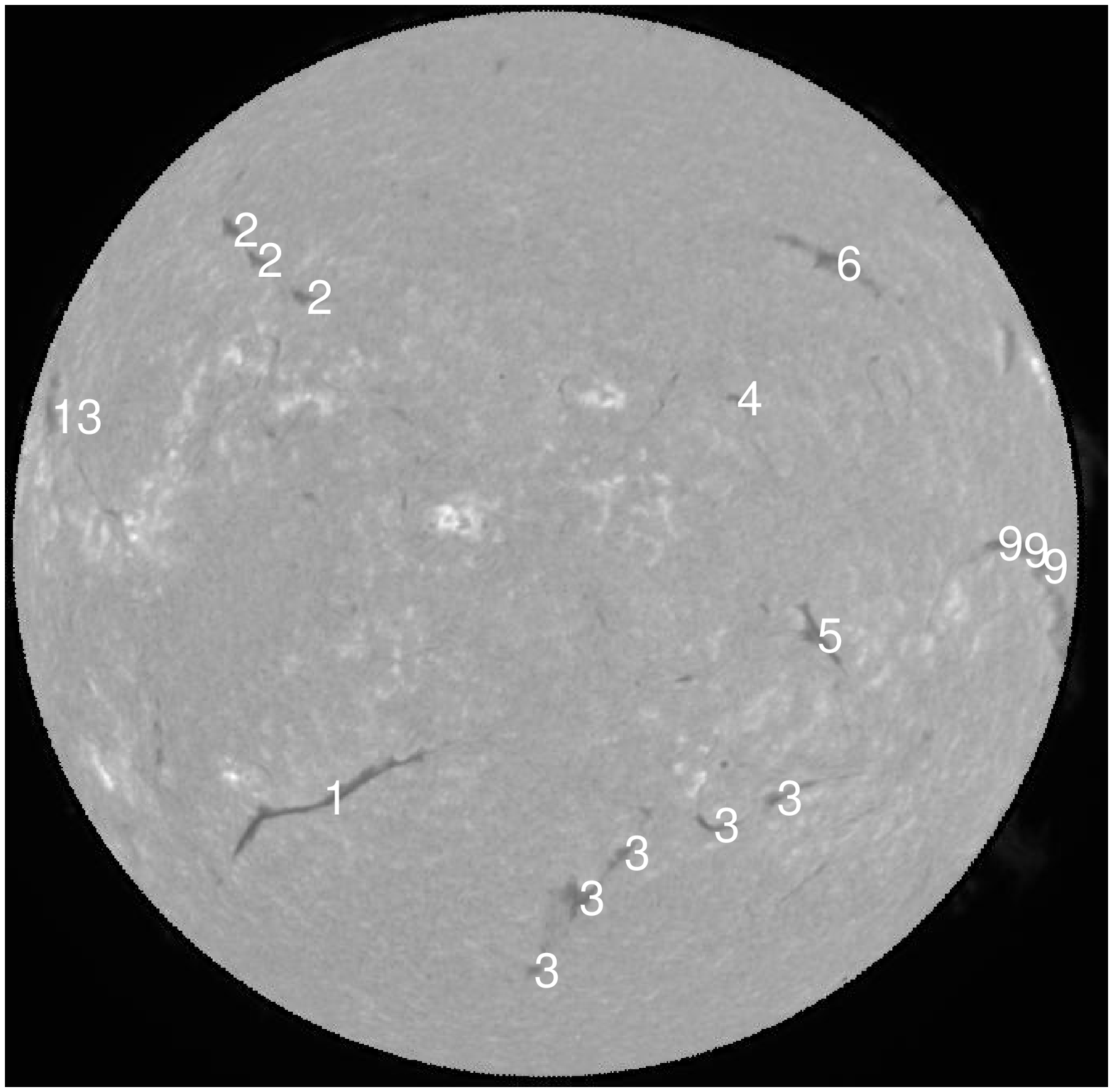}
              }
     \vspace{-0.091\textwidth}
     \centerline{\Large \bf
      \hspace{-0.01 \textwidth} \color{white}{(c)}
      \hspace{0.44\textwidth}  \color{white}{(d)}
         \hfill}
     \vspace{0.02\textwidth}
\caption{
An example of filament tracing method between two images at 18:41:38 UT, 17 February 2001 and at 18:41:40 UT, 18 February 2001 respectively obtained by MLSO. (a) The detected result of the earlier image, each filament is labeled with a unique number. (b)the detected filament fragments are shown individually;
(c) the updated filaments where fragments are labelled to be one filament.
(d) Final result, where the labels of the traced filaments are updated.
        }
   \label{mlso_trace}
   \end{figure}

\section{Filament tracing}
\label{Tracing}

Tracing the evolution of the filament is important for understanding
the physical nature and the solar cyclic variation of the filaments.
In this section we present a tracing method. Here we use the filament
label, position and area as the input parameters, which have been obtained
in Section~\ref{Feature}, to trace the daily evolution of the filaments.

We define two input images as $I_{\rm old}$ (\textit{i.e.} the image observed
at the old time) and $I_{\rm new}$ (\textit{i.e.} the image at the new time).
The main steps of the tracing method are as follows:

\textbf{i)} We obtain the observation time of the two images $I_{\rm old}$ and
$I_{\rm new}$, then calculate the time interval $T_{\rm interval}$;

\textbf{ii)} Using the latitude from the position features of each filament in
$I_{\rm old}$ (or $I_{\rm new}$) in order to calculate the rotation velocity at
this latitude $\omega_{\rm old}$ (or $\omega_{\rm new}$), then calculate the
possible longitude $PLO_{\rm old}$ (or $PLO_{\rm new}$) with the time interval
$T_{\rm interval}$. Here the $PLO_{\rm new}$ is calculated by assuming that the
Sun rotates backward;

   \begin{equation}
   PLO_{\rm old(\textit{i})} = CLO_{\rm old(\textit{i})} + \omega_{\rm old(\textit{i})}\cdot T_{\rm interval} \,
   (i = 1, 2,\ldots, n) \, ,
   \end{equation}

    \begin{equation}
   PLO_{\rm new(\textit{j})}= CLO_{\rm new(\textit{j})} - \omega_{\rm new(\textit{j})}\cdot T_{\rm interval} \,
   (j = 1, 2,\ldots, n) \, ,
   \end{equation}
\noindent
where $CLO_{\rm old(i)}$ (or $CLO_{\rm new(j)}$) is the current longitude of
$i$-th (or $j$-th) filament in $I_{\rm old}$ (or $I_{\rm new}$). Here, we adopt
the solar rotation angular velocity formula \cite{1986Balthasar} to
determine $\omega_{\rm old(i)}$ and $\omega_{\rm new(j)}$:

   \begin{equation}
   \omega(\theta)=(14.551\pm0.006)-(2.87\pm0.06)\sin^{2}\theta \, ,
   \end{equation}

\noindent
where $\omega$ is the angular velocity (degrees per day) and $\theta$
the latitude;

\textbf{iii)} We obtain the possible position [$PP_{\rm old}: (PLA_{\rm old}, PLO_{\rm old})$] of the
filament after $T_{\rm interval}$ via the differential rotation formula, then
calculate the distance $D_{\rm old\rightarrow new}$ between
$PP_{\rm old}$ and real current position [$CP_{\rm new}: (CLA_{\rm new}, CLO_{\rm new})$]
of the filament in $I_{\rm new}$. Because the drift velocity of the filament
is much smaller than the solar rotation velocity, we assume the
filament latitude does not change in $I_{\rm old}$ (\textit{i.e.} $PLA_{\rm old}=CLA_{\rm old}$,
$PLA_{\rm old}$ is the possible latitude and $CLA_{\rm old}$ is the current
latitude in $I_{\rm old}$). The distance $D_{\rm old\rightarrow new}$ between
$PP_{\rm old}$ and $CP_{\rm new}$ is:

    \begin{equation}
   D_{\rm old{\rightarrow}new}=\sqrt{(PLA_{\rm old}-CLA_{\rm new})^{2}+(PLO_{\rm old}-CLO_{\rm new})^{2}}\,.
   \end{equation}

\textbf{iv)} We assume that the Sun rotate backward, then obtain the possible position of the
filament [$PP_{\rm new}: (PLA_{\rm new}, PLO_{\rm new})$] after $T_{\rm interval}$. A processing
which is similar to 3) is processed, the distance
$D_{\rm new{\rightarrow}old}$ between $PP_{\rm new}$ and $CP_{\rm old}$ is:

    \begin{equation}
   D_{\rm new{\rightarrow}old}=\sqrt{(PLA_{\rm new}-CLA_{\rm old})^{2}+(PLO_{\rm new}-CLO_{\rm old})^{2}}\,.
   \end{equation}

\textbf{v)} For each filament in $I_{\rm new}$, we check all of the
filaments in $I_{\rm old}$. Only if one meets the following three
conditions, the filament in $I_{\rm old}$ would be considered to be the same
filament and marked with the same label as in $I_{\rm new}$:

\textbf{Condition 1:} $D_{\rm old{\rightarrow}new}\leq{ED_{1}}\cdot
T_{\rm interval}$;

\textbf{Condition 2:} $D_{\rm new{\rightarrow}old}\leq{ED_{2}}\cdot
T_{\rm interval}$;

\textbf{Condition 3:} $Area_{\rm old(\textit{current})}/Area_{\rm new(\textit{j})}\geq{ratio}, (j=1, 2,\ldots, n),$

\noindent
where $ED_{1}\cdot T_{\rm interval}$ and $ED_{2}\cdot T_{\rm interval}$ are the distance
threshold. In our program we take $ED_{1}=ED_{2}=100$ pixels day$^{-1}$.
$Area_{\rm old(\textit{current})}$ and $Area_{\rm new(\textit{j})}$ are the filament
area at the old and new times, respectively. Condition 3
gives the maximum proportion of the deformation, \textit{i.e.} less than the specified
$ratio$ of the size of the previous filament. Here we set $ratio=50\%$ . If the
three conditions are not satisfied, the filament at the new time would be
identified as a new filament and given a new label. This step continues
until all filaments in $I_{\rm new}$ are treated. Finally, the
filament labels in $I_{\rm new}$ are updated. Figure~\ref{mlso_trace}
gives an example of our tracing method, where panel (a) shows
the detected result of the earlier image with each filament labeled a
unique number, panel (b) depicts the filament fragments after detection,
penel (c) shows the updated filaments where several fragments merged into one
filament, and panel (d) shows the tracing result based on the
earlier image in panel (a). For example, the filament 2 in the earlier image (panel a) is split into three fragments (labeled 2, 4, and 5 in the later
image (panel b). After using the tracing method, they were labeled the same
number in the earlier image (panel a). Filament 13 in the later image
(panel d) was not detected in the earlier image (panel a), so it was given a
new number.

\section{Performance}
\label{Performance}

Our code was developed by using $\rm MATLAB^{\tiny \circledR}$ Desktop Tools
and Development Environment on a desktop computer (CPU: $\rm Intel^{ \tiny
\circledR}\ \rm Core^{\tiny TM}$ Duo 3.00 GHz). After processing of each image
file, the result (such as the label, the position, the area, and other
features) are written to a text file. The average processing speed is 1 second
for the filament detection in a single image and 3.5 seconds for the filament
detection and tracing in two images. We randomly selected 100 images from the
MLSO H$\alpha$ archive for testing, and compared the automated result with the manual ones. For filaments and filament fragments, the two methods are
overlapping by $85\pm2\%$ and $88\pm4\%$, respectively. The error includes two
types of false recognition: One is that there is a manually recognized
filament, but the
automated method cannot detect it. The other is that the automated method
detects a filament, but it does not appear on the real solar disk. It is noted
that the latter is rarely seen in our method. If a filament splits into several small fragments, and the criterion of the ratio of the long to short axes is not
satisfied (\textit{i.e.} being recognized as a sunspot), our method may miss these fragments. The accuracy of the filament fragment number is a little
higher than that of the filament, which is due to the prescribed ``distance criterion". Sometimes several filaments or filament fragments in one active region are so close to each other and within the ``distance criterion", they are recognized as one filament. This kind of false recognition does appear in our filament fragment detection and we have to improve the filament fragments merging method in the future. For other features such as position, perimeter,
area, and spine, there are no standard criteria to test the accuracy of the results processed by our codes. However, we defined two indices in order to test the performance of our method, \textit{i.e.} the ``edge closed rate" and the ``area fully filled rate". The ``edge closed rate" is defined as the number of detected filaments with edge curve closed as a percentage of the total number of detected filaments among the 100 test images. We found that the rate is 91\%. This rate mainly depends on the selection of the threshold in the threshold filter
processing and the two thresholds in the Canny edge-detection method. If the
filament edge curve is not closed, it leads to low detection accuracy and
affects the subsequent processing. The ``area fully filled rate" is defined as
the number of the detected filaments with edge curve fully filled with foreground pixels as a percentage of the total number of detected filaments among the 100 test images, and the rate is 75\%. After the edge detection is finished, if the edge curve is not closed, the morphological object filling method could not fully fill in the area enclosed by the edge curve. This leads to the decrease in ``area fully filled rate" and the detection accuracy rate.
The filament spine is also affected by the area problem, \textit{i.e.} if the area is not fully filled, our method may get a wrong topology of the filament after the morphological skeletonization processing. Furthermore, if the barbs are located near the end of the filament spine or the filament size is
relatively small, the recognized spine may be shorter than the real length
after the morphological barb removal processing.
The shorter the time interval is, the higher the tracing
accuracy is. Here, we set the default time interval to be about one day,
the accuracy of the tracing method is about 80\%. In addition, we
also test images from Big Bear Solar Observatory (BBSO) H$\alpha$ archive
(\url{ftp://ftp.bbso.njit.edu/pub/archive}) to validate the versatility of
our method. The results are similar and satisfactory.

\section{Statistical Results of the Filament Latitude}
\label{Results}

We use our automated method to analyze 3470 images obtained by MLSO from
January 1998 to December 2009. In this section, we present the statistical
results of the evolution of the filament latitudinal distribution because of
its relatively high accuracy. Furthermore, from a statistical point of view,
such results can be significant in understanding the cyclic migration of
solar filaments.

\subsection{Butterfly Diagrams}

\begin{figure}    
\centerline{\hspace*{-0.05\textwidth}
            \includegraphics[width=1.0\textwidth,clip=]{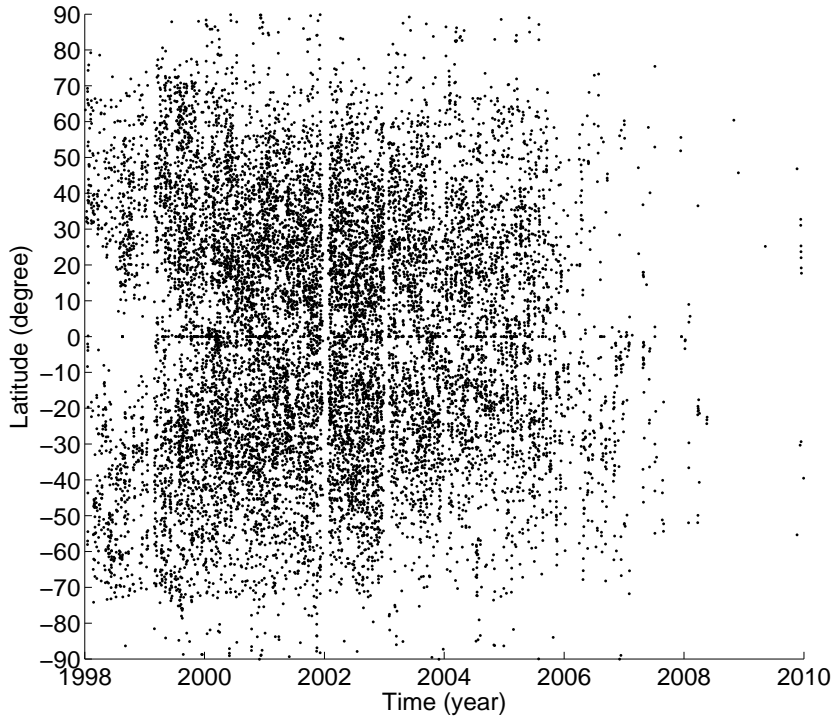}}
\caption{Butterfly diagram of filaments from January 1998 to December 2009 in the Solar Cycle 23. Each dot represents a single observation.}
\label{BFD_f}
\end{figure}

For the period from January 1998 to December 2009, we process one image per day and have detected
13,832 filaments. The temporal evolution of the latitudinal distribution of these
filaments is depicted as the scatter plot in Figure~\ref{BFD_f} (each dot represents a single observation), where we can
clearly see a butterfly diagram, similar to sunspots. Because of the lack
of observations in some periods, there are several white vertical gaps in the
butterfly diagram.  From the diagram we can see the distribution and
the migration of the filaments. This butterfly diagram indicates that the
formation of the filaments mainly migrates towards the equator from the beginning to the
end of the Solar Cycle 23.

\subsection{Drifting Velocity}

\begin{figure}    
\centerline{\hspace*{-0.05\textwidth}
            \includegraphics[width=1.0\textwidth,clip=]{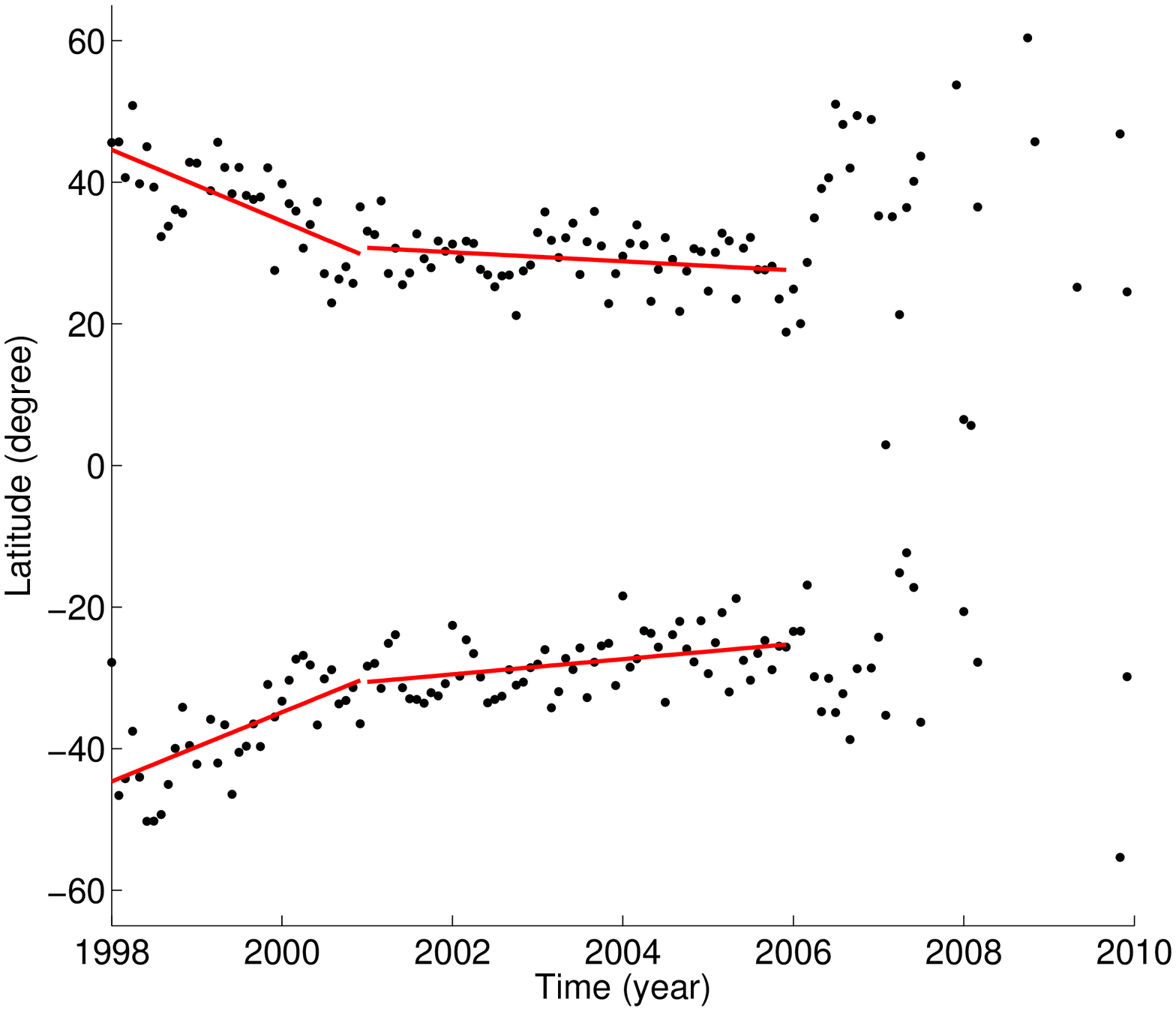}}
\caption{Temporal evolution of the monthly mean latitudinal distribution of
filaments from January 1998 to December 2009. Linear fittings are shown by solid-red lines.
}
\label{mm_total}
\end{figure}

\begin{figure}    
\centerline{\hspace*{-0.05\textwidth}
            \includegraphics[width=1.0\textwidth,clip=]{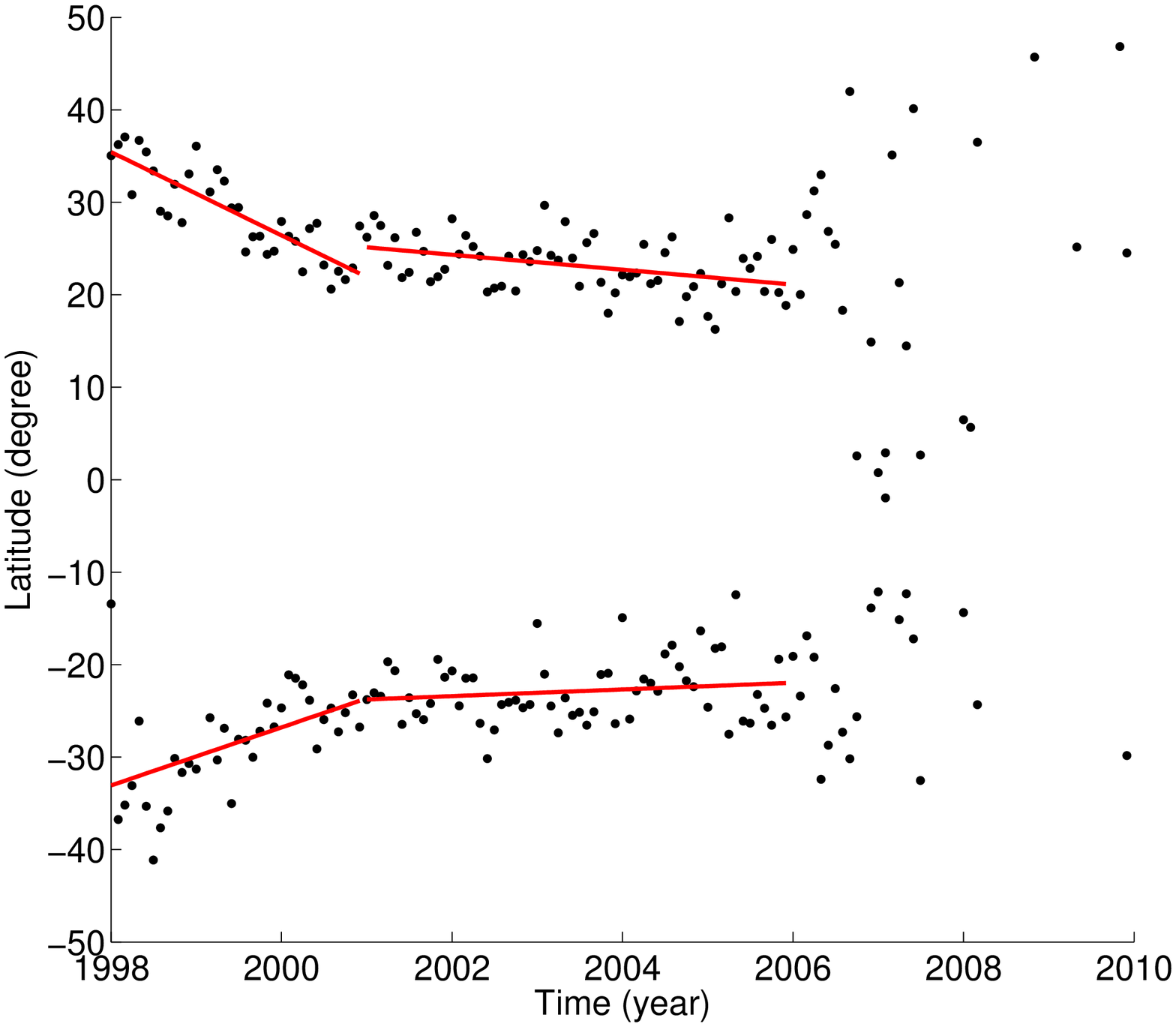}}
\caption{Temporal evolution of the monthly mean latitudinal distribution
of filaments with a latitude lower than $50^{\circ}$ from January 1998 to December 2009. Linear fittings are shown by solid-red lines. }
\label{mm_0-50}
\end{figure}

\begin{figure}    
\centerline{\hspace*{-0.05\textwidth}
        \includegraphics[width=0.9\textwidth,clip=]{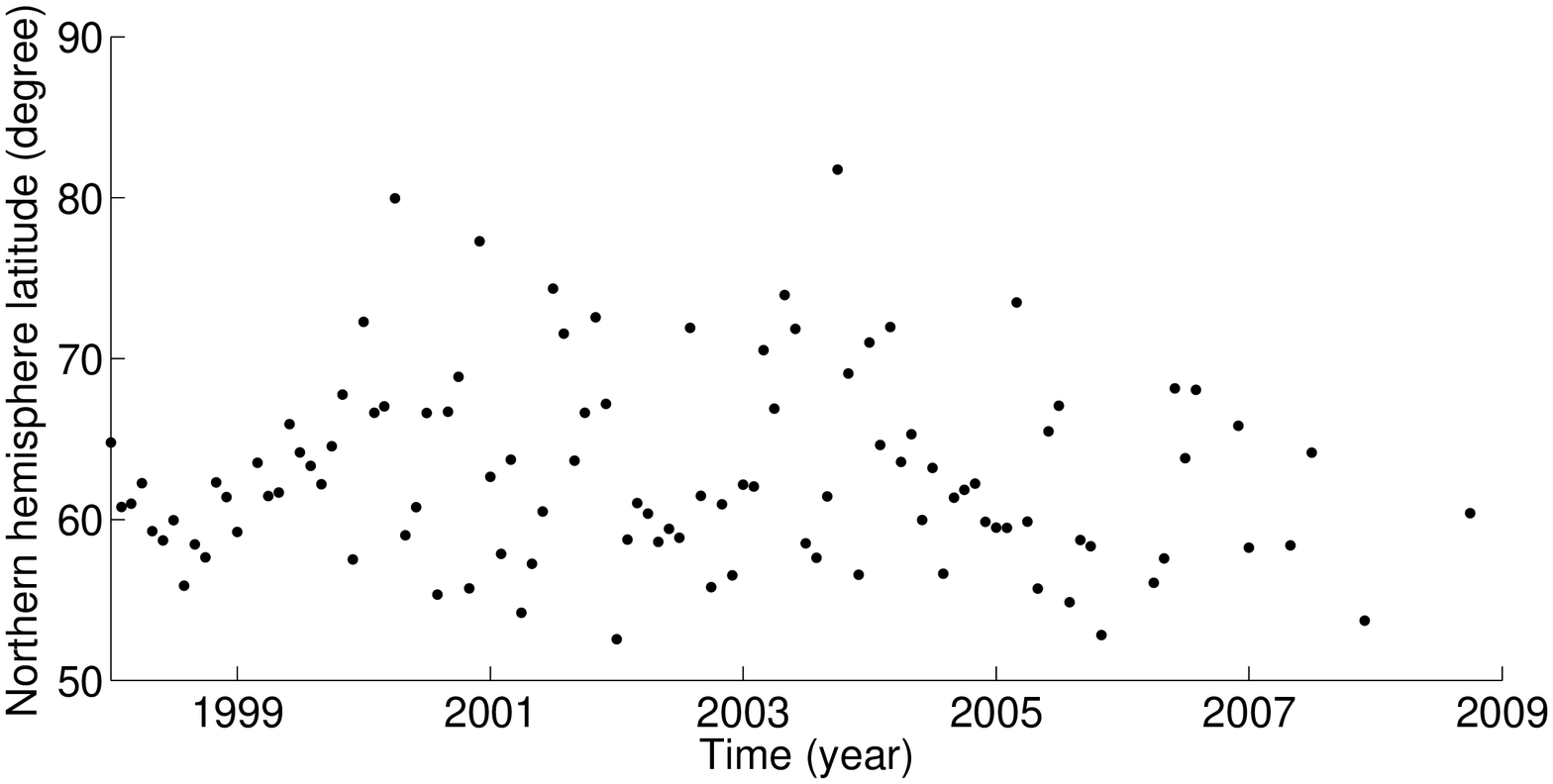}
        }
     \vspace{0.0\textwidth}
\centerline{\hspace*{-0.05\textwidth}
        \includegraphics[width=0.9\textwidth,clip=]{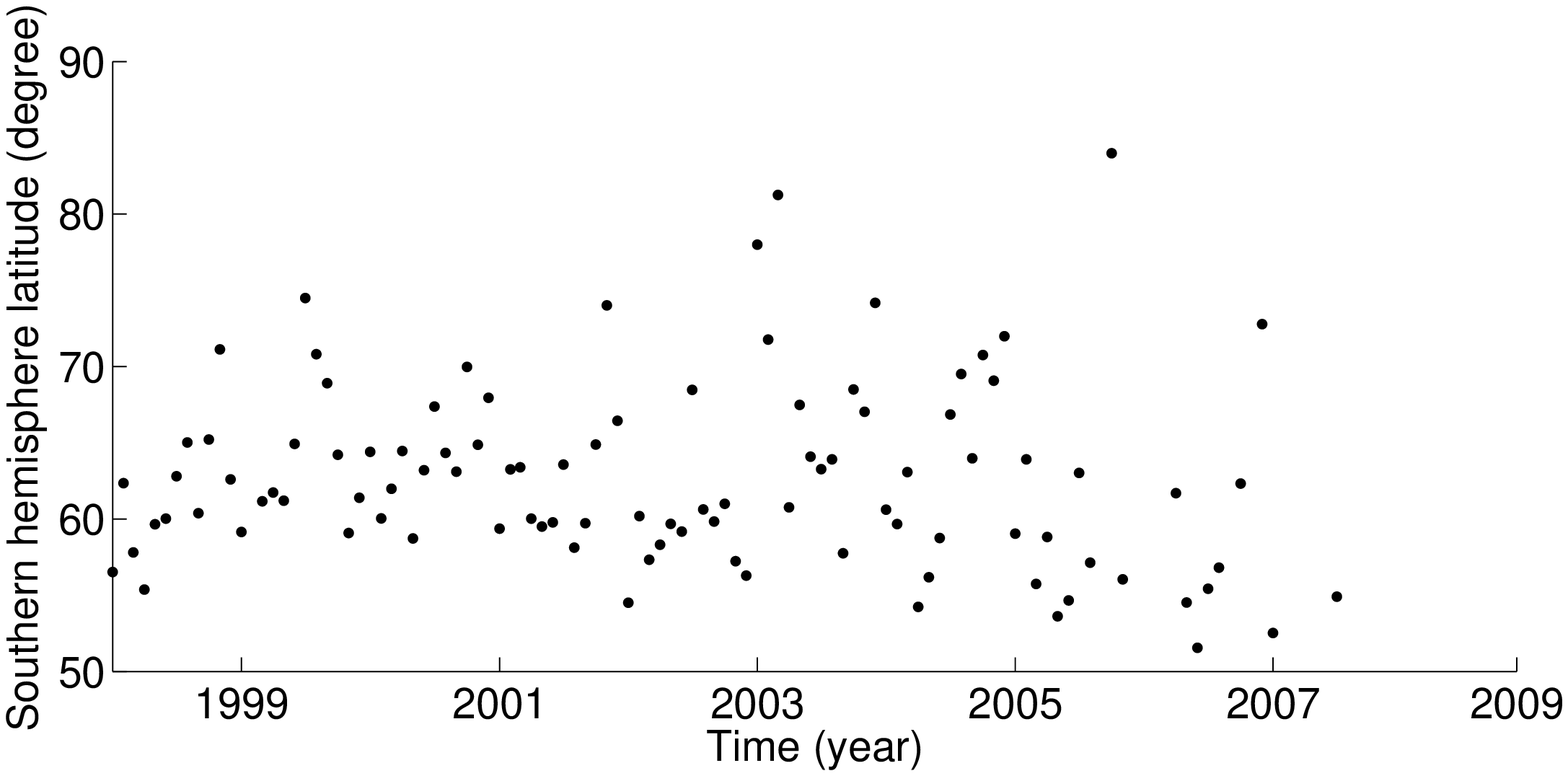}
        }
    \vspace{0.0\textwidth}
\caption{Temporal evolution of the monthly mean latitudinal distribution of
the filaments with latitude higher than $50^{\circ}$ from January 1998 to December 2009. The
upper panel is for the northern hemisphere and the bottom panel is for the
southern hemisphere.}
\label{mm_50-90}
\end{figure}

From the butterfly diagram we only get qualitative results, as mentioned by
\inlinecite{2010Li}. In order to make a quantitative analysis, we adopt the
monthly mean latitude of the filaments in the northern and southern
hemispheres, respectively. The calculated results are plotted in Figure
\ref{mm_total}. It can be seen that the monthly mean latitudinal distribution
of the filaments has three drift trends: from 1998 to the solar maximum (2001)
the drift velocity is very fast. After the solar maximum it becomes relatively
slow. After 2006, the drift velocity becomes divergent. A linear fitting is
used to the data points in different periods, resulting in an average drift
velocity being 0.0138 degree day$^{-1}$, or 1.86 m s$^{-1}$, during
1998--2001, and 0.0017 degree day$^{-1}$ or 0.23 m s$^{-1}$ during 2002--2006
in the northern hemisphere. It is 0.0134 degree day$^{-1}$, or 1.80 m s$^{-1}$
during 1998--2001 and 0.0029 degree day$^{-1}$, or 0.39 m s$^{-1}$ during
2002--2006 in the southern hemisphere. Here, we did not fit the monthly
mean filament distribution after 2006, because it becomes divergent near the
solar minimum.

Since the normal solar activity is usually applied to the events with
latitudes lower than $50^{\circ}$ \cite{1998Sakurai,2008Li}, we analyze the
filaments with latitudes lower than $50^{\circ}$. The calculated result is
plotted in Figure~\ref{mm_0-50}. It can be seen that the monthly mean
latitudinal distribution of these filaments again has three drift trends: From
1998 to the solar maximum (2001) the drift velocity is fast, \textit{i.e.} 0.0123
degree day$^{-1}$ or 1.66 m s$^{-1}$ in the northern hemisphere and 0.086 degree day$^{-1}$
or 1.16 m s$^{-1}$ in the southern hemisphere. After the solar maximum the drift
velocity becomes relatively slow, \textit{i.e.} 0.0022 degree day$^{-1}$ or 0.29 m s$^{-1}$ in
the northern hemisphere and 0.0010 degree day$^{-1}$ or 0.13 m s$^{-1}$ in the southern
hemisphere, respectively. After 2006 it becomes divergent. These results are
similar to those of the entire latitudinal distribution. The reason is easy to
understand: among the 13,832 filaments we detected, only 1,130 filaments have
latitudes higher than $50^{\circ}$. In other words, the detected filaments are
mainly distributed in latitudes lower than $50^{\circ}$. There is no obvious difference
between the northern and the southern hemispheres. These results are similar to
the statistical results of \inlinecite{2010Li}. Similarly, we plot the monthly
mean latitude of the filaments with latitudes higher than $50^{\circ}$ in
Figure~\ref{mm_50-90}. However, no clear trend is discernible.

  \begin{figure}    
   \centerline{\hspace*{0.015\textwidth}
               \includegraphics[width=0.515\textwidth,clip=]{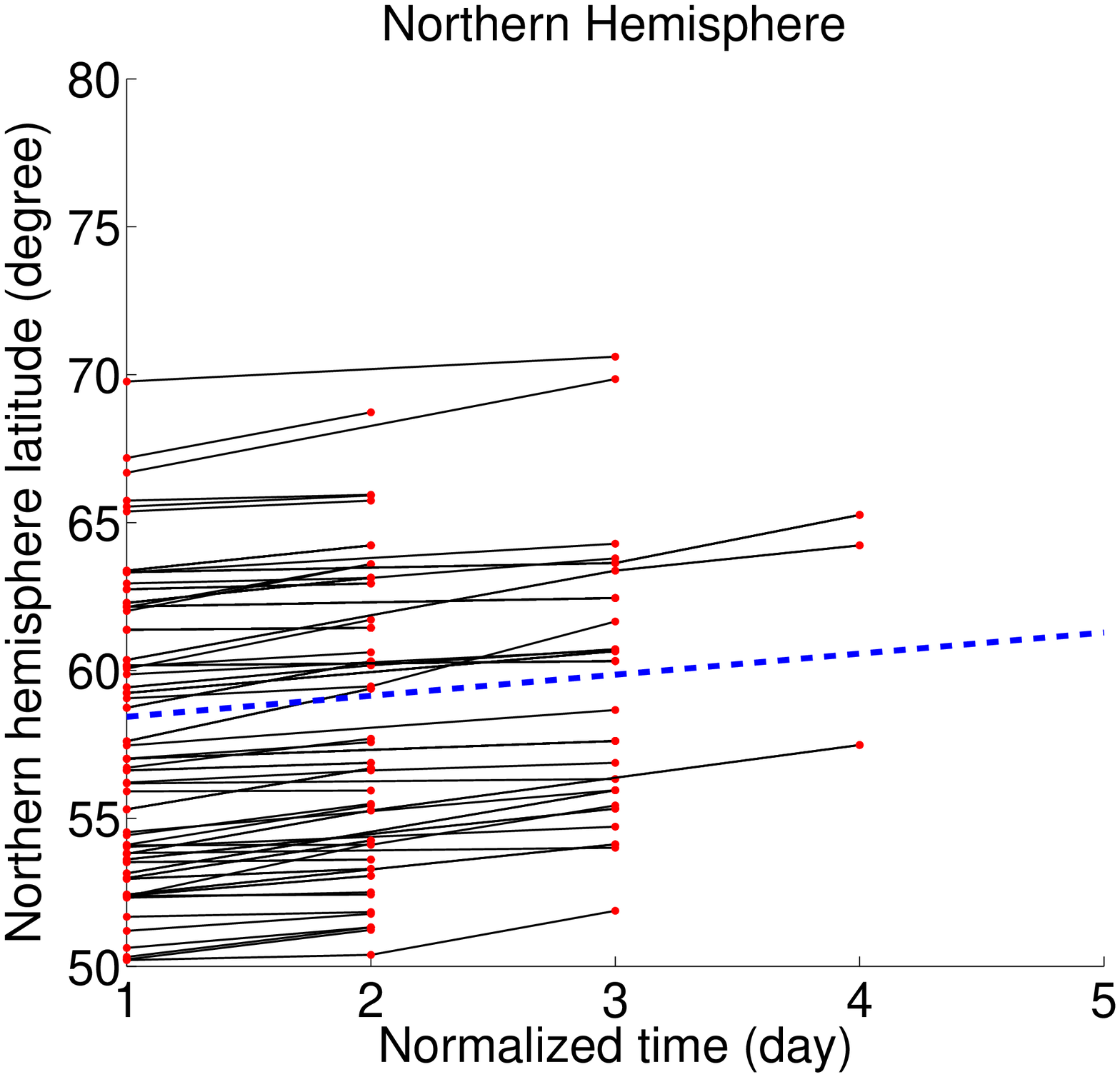}
               \hspace*{-0.03\textwidth}
               \includegraphics[width=0.515\textwidth,clip=]{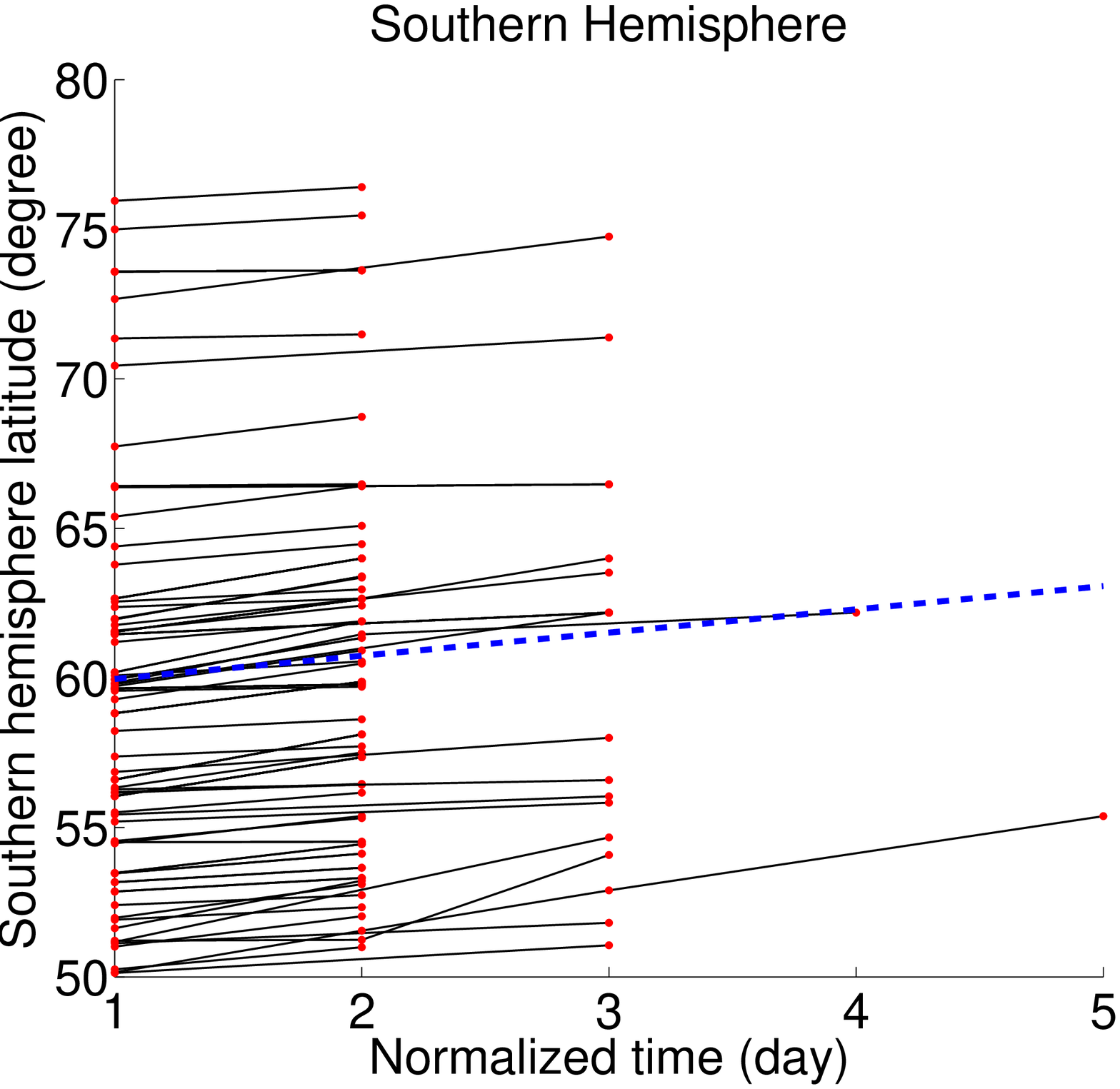}
              }
   \vspace{0.01\textwidth}
\caption{Traced filament latitude variation \textit{versus} normalized time. Here the normalized time means that we put the dates of the first detection of the filaments as the start time.The left
and right panels are for the northern and southern hemisphere, respectively.
The solid-black line represents the temporal and
spatial variation of the traced filament latitudes, and the red dot
indicates the filament where and when it was detected and traced.
The dash-blue line is the linear fitted average of all traced filament
temporal and spatial variations. }
\label{normalized}
\end{figure}

\begin{figure}    
\centerline{\hspace*{0.015\textwidth}
            \includegraphics[width=0.515\textwidth,clip=]{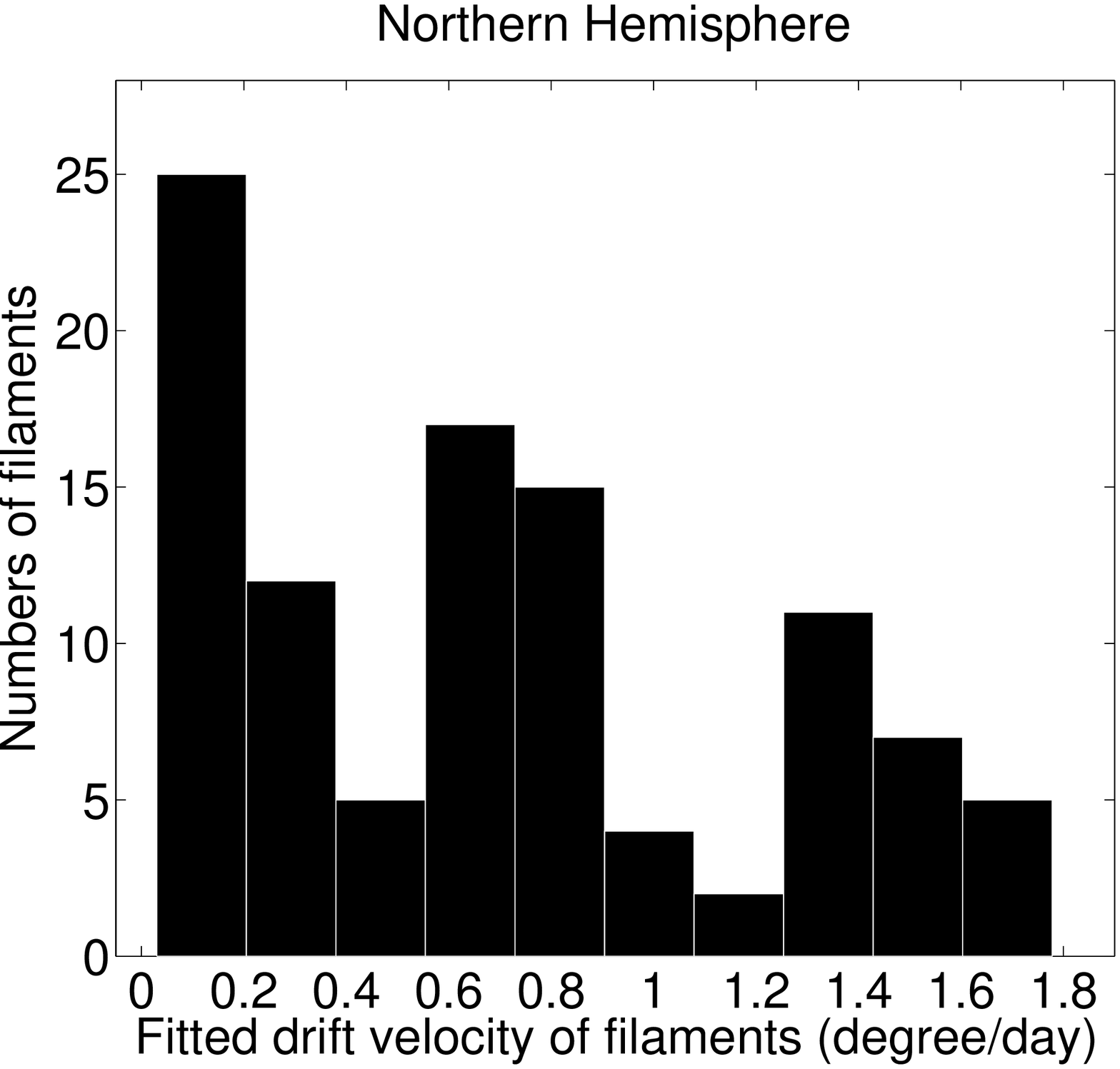}
            \hspace*{-0.03\textwidth}
            \includegraphics[width=0.515\textwidth,clip=]{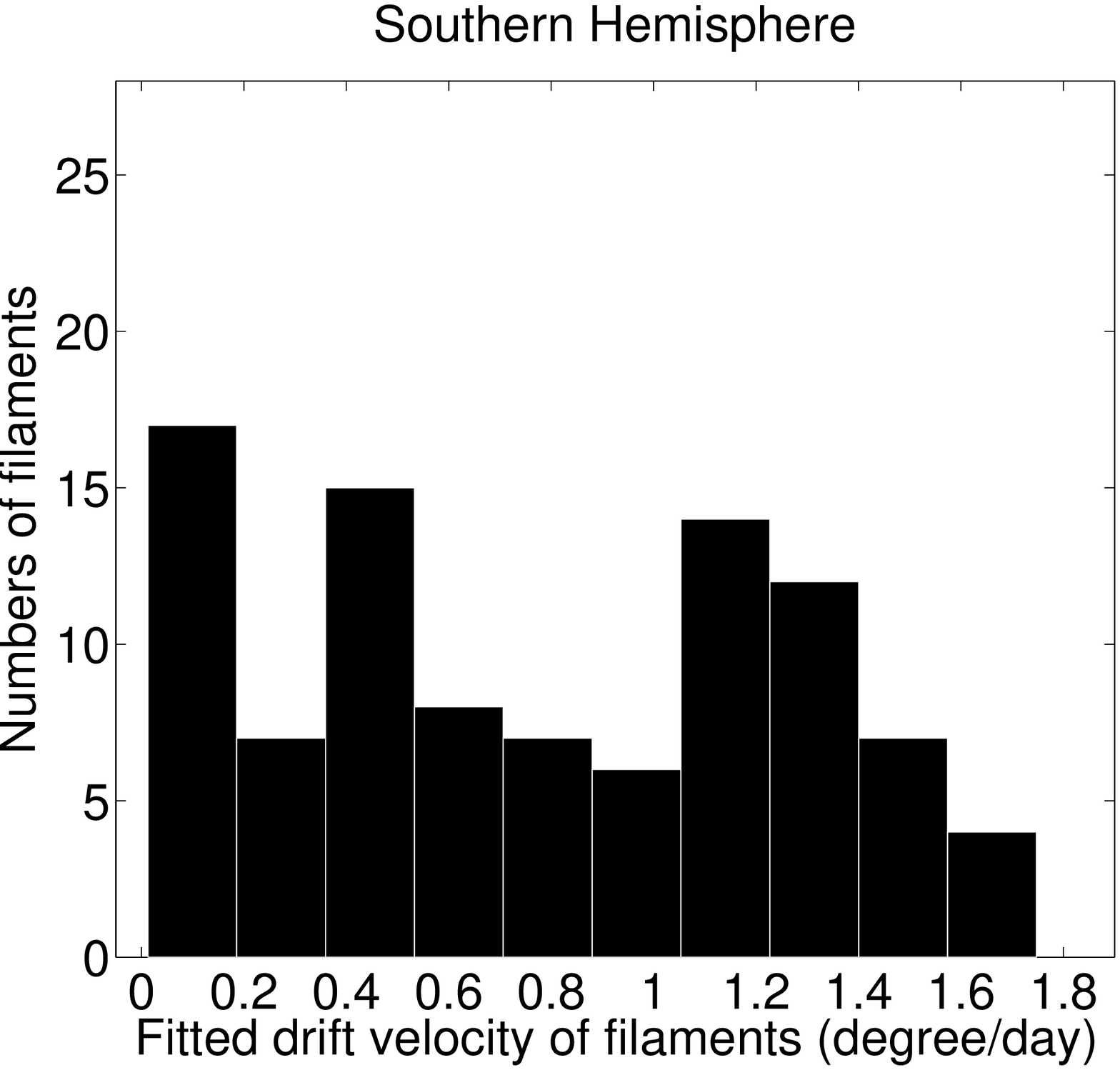}
             }
\vspace{0.01\textwidth}
\caption{Histograms of the drift velocity distribution of the filament whose
latitudes are higher than $50^{\circ}$. The left panel is for the northern
hemisphere with 103 filaments in total and the right panel for
the southern hemisphere with 97 filaments in total.
} \label{hist}
\end{figure}

In order to find the migration of the individual filaments above $50^{\circ}$,
we employ our tracing method and set three additional conditions for tracing:

\textbf{Condition 1:} The filament positions are higher than
$50^{\circ}$ at the first detection;

\textbf{Condition 2:} The time interval is less than two days. In other words,
in three consecutive days, observations are available in at least two days. The
purpose of this condition is to improve the accuracy.

\textbf{Condition 3:} The total time lapse should be less than ten
days, because one specific filament observation can be clearly visible in
less than a half-period of the solar rotation. If the time lapse is greater,
the accuracy of the tracing method is lower.

We plot the tracing results of the latitude versus normalized time in Figure
\ref{normalized}. Here the normalized time means that we put the dates of the
first detection of the filaments as the start time. The drift velocity
distribution histograms in the northern hemisphere and the southern
hemisphere are shown in Figure~\ref{hist}. In the northern hemisphere, there
are 103 filaments (which occupy 57\% of all filaments satisfying the conditions
and being traced with the latitude higher than $50^{\circ}$), which migrate
towards the polar region. The average drift velocity is 0.7126
degree day$^{-1}$ (96.2 m s$^{-1}$). In the southern hemisphere, there are 97 filaments (which occupy 61\% of all filaments satisfying the conditions and
being traced with latitude higher than $50^{\circ}$), which migrate towards the polar region. The average drift velocity is 0.7771
degree day$^{-1}$ (104.9 m s$^{-1}$). From Figure~\ref{hist}, we found that the
drift velocities of the filaments with latitudes higher than
$50^{\circ}$ are divergent, while most of these filaments migrate towards the polar
region with relatively high velocities. Such a result is similar to that
of \inlinecite{1982Topka}. However, they found that the poleward drift velocity
is about 10 m s$^{-1}$, which is much slower than ours.

\section{Conclusions}
\label{Conclusion}
We have developed a method to automatically detect and trace solar filaments
from H$\alpha$ full-disk images. The program consists of three parts: First, a
preprocessing module is applied to correct the original images. Top-hat
enhancement enables us to clearly distinguish the filaments from non-filament
features. Second, we introduce the Canny edge-detection method to segment and
detect filaments. This method gives us a precise filament edge. Third, our
program routines recognize filament features through the morphological
operators. We randomly
selected 100 images from MLSO observations to test our method, which
is demonstrated to be robust and efficient. For the filament detection, the similarity between the machine recognition and human vision is $85\pm2\%$.
The solar rotation, the filament position, and the deformation of the filament
have been considered in order to trace the filament evolution. The accuracy of
the tracing method is about 80\% when the time interval is about one day. In
addition, our program can process images not only in different file formats, but
also from different observatories.

We used our method to automatically process and analyze 3470 images obtained by
MLSO from January 1998 to December 2009 . A butterfly diagram of filaments is
obtained, where we can clearly see that filaments move mainly towards the
equator in both hemispheres. In order to obtain more quantitative results, we
calculated the monthly mean latitudes of the filaments whose latitudes are
within $0^{\circ}-50^{\circ}$ or higher than $50^{\circ}$ in both northern and
southern hemispheres, respectively. Furthermore, we use our tracing method to
trace the evolution of the individual filaments with a latitude higher than
$50^{\circ}$. Our main conclusions are listed as follows:

$\bullet$ The latitudinal migration of solar filaments have three trends in the
Solar Cycle 23: from 1998 to 2001 (the solar maximum) the drift velocity is
fast. From the solar maximum to the year 2006 the drift velocity becomes
relatively slow. After 2006, \textit{i.e.} near solar minimum, the migration
becomes divergent.

$\bullet$ About 60\% filaments with latitudes higher than $50^{\circ}$
migrate towards the polar region with relatively high velocities in
both northern and southern hemispheres.

$\bullet$ The difference of the latitude migration of the filaments
between the northern and southern hemispheres is not obvious in the
Solar Cycle 23.

We will improve our method to be more reliable and efficient, and apply it to
the observational data from our \textit{Optical \& Near Infrared Solar Eruption
Tracer} (ONSET) in Nanjing University \cite{Fang2012}.

%

%
\begin{acks}
The authors thank the Mauna Loa Solar Observatory team for making the data
available and Sun J. Q. for his help in identifying filaments. We also thank
the referee very much for the constructive suggestions
which greatly improved the paper in various ways.
This work is supported by the National Natural Science Foundation of
China (NSFC) under the grants 10221001, 10878002, 10403003, 10620150099,
10610099, 10933003, 11025314, and 10673004, as well as the grant from the
973 project 2011CB811402.
\end{acks}


\end{article}
\end{document}